\documentclass{statsmsc}

\title{GNAR-HARX Models for Realised Volatility: Incorporating Exogenous Predictors and Network Effects}
\author{Tom Ó Nualláin}
\CID{06021498}
\supervisor{Professor Guy Nason}
\date{\today}
\logoimg{}

\usepackage{xcolor}
\usepackage{tikz}
\usetikzlibrary{positioning, arrows.meta, shapes.multipart}
\usepackage{hyperref}
\usepackage{booktabs}
\hypersetup{
    colorlinks,
    linkcolor={blue!50!black},
    citecolor={blue!50!black},
    urlcolor={blue!50!black}
}
\usepackage{graphicx}
\usepackage{float}

\begin{document}

\maketitle

\declarationname{Tom Ó Nualláin}
\declarationdate{\today}
\declaration 

\begin{acknowledgements}
I would first like to express my sincere gratitude to my supervisor, Professor Guy Nason, for his generous guidance, encouragement, and advice throughout the summer. His support has been invaluable in shaping this project and in developing my understanding of the topic.

I am also grateful to Dr. Mikko Pakkanen for providing access to the main dataset used in this study. I would like to thank Peiyi Zhou for the shared discussions during our joint supervision meetings, which were both motivating and helpful.

Special thanks go to Marcos Tapia Costa, who generously shared code that informed parts of my implementation, and to Henry Palasciano, whose prior code served as a useful reference in developing my own.

Finally, I wish to thank my family and friends for their constant support and encouragement throughout the summer.
\end{acknowledgements}

\mainmatter

\begin{abstract}
This project introduces the GNAR-HARX model, which combines Generalised Network Autoregressive (GNAR) structure with Heterogeneous Autoregressive (HAR) dynamics and exogenous predictors such as implied volatility. The model is designed for forecasting realised volatility by capturing both temporal persistence and cross-sectional spillovers in financial markets. We apply it to daily realised variance data for ten international stock indices, generating one-step-ahead forecasts in a rolling window over an out-of-sample period of approximately 16 years (2005–2020).

Forecast accuracy is evaluated using the Quasi-Likelihood (QLIKE) loss and mean squared error (MSE), and we compare global, standard, and local variants across different network structures and exogenous specifications. The best model found by QLIKE is a local GNAR-HAR without exogenous variables, while the lowest MSE is achieved by a standard GNAR-HARX with implied volatility. Fully connected networks consistently outperform dynamically estimated graphical lasso networks.

Overall, local and standard GNAR-HAR(X) models deliver the strongest forecasts, though at the cost of more parameters than the parsimonious global variant, which nevertheless remains competitive. Across all cases, GNAR-HAR(X) models outperform univariate HAR(X) benchmarks, which often require more parameters than the GNAR-based specifications. While the top model found by QLIKE does not use exogenous variables, implied volatility and overnight returns emerge as the most useful predictors when included. 
\end{abstract}

\section{Introduction}
\label{sec:introduction}

Given its central role in portfolio allocation, risk management, derivatives pricing, and algorithmic trading, the accurate modelling of realised volatility has long been an active area of research. Accurate forecasts can enable market participants to hedge risk, manage uncertainty, and optimise capital allocation. This has led to an extensive literature on statistical methods for volatility modelling, increasingly informed by the availability of high-frequency financial data.

An important empirical feature of realised volatility is its long memory, evidenced by a slowly decaying autocorrelation function. Traditional ARMA models struggle to capture this persistence unless made high-order, leading to increased estimation complexity. While ARFIMA models \citep{andersen_modeling_2003} address this using fractional differencing, they are challenging to estimate in practice \citep{baillie_long_1996} and may provide less direct financial interpretation compared to other models.

One widely studied approach is the Heterogeneous Autoregressive (HAR) model introduced by \cite{corsi_simple_2009}, which approximates long memory using a small number of lags at daily, weekly, and monthly horizons. Motivated by the idea that market participants operate at different time scales, HAR models capture heterogeneity in volatility dynamics while remaining straightforward to estimate. As summarised by \cite{clements_practical_2021}, the HAR model and its variants consistently deliver competitive, and often superior, forecast performance across a wide range of assets and market conditions.

While HAR models effectively capture persistence in univariate volatility, they are confined to individual assets or markets, and therefore cannot address cross-market spillovers. Recognising this limitation has motivated the development of multivariate approaches that explicitly model interconnected markets.

Recent developments in financial econometrics increasingly emphasise the interconnected nature of global markets. Volatility shocks often propagate across borders, particularly during turbulent periods, a phenomenon widely known as volatility spillover \citep{rigobon_spillovers_2003}. For example, \citet{buncic_global_2016} and \citet{wang_volatility_2018} show that U.S. market volatility influences foreign asset markets, with the VIX frequently acting as a global risk barometer.

Traditional multivariate approaches such as Vector Autoregressive (VAR) models, BEKK-GARCH \citep{engle_multivariate_1995}, and Wishart autoregressions \citep{gourieroux_wishart_2009} can capture cross-market dynamics but scale poorly: parameter counts grow rapidly with the number of assets, leading to increased estimation variance and reduced predictive power \citep{callot_modeling_2017}. This curse of dimensionality has motivated the search for more parsimonious yet flexible approaches to capturing volatility spillovers, an objective for which network-based time series models are particularly well suited.

In this context, network time series models provide a promising framework by capturing both temporal and cross-sectional dependencies in multivariate time series through an underlying network representation.

An early contribution in this area is \cite{knight_modelling_2016}, who introduced Network Autoregressive Moving Average (NARIMA) processes. These models extend the VARMA framework to network settings and are particularly suited to cases where the network structure itself may evolve over time.

\cite{zhu_network_2017} then proposed the Network Autoregressive (NAR) model, where each node evolves as a function of its own lags and those of its immediate (one-hop) neighbours. The NAR model reduces complexity relative to VAR by restricting dependence to local interactions rather than requiring interactions across all pairs of nodes. However, the NAR model assumes a binary, static network and homogeneous dynamics across nodes.

Building on \citet{knight_modelling_2016}, the Generalised Network Autoregressive (GNAR) model of \citet{knight_generalised_2019, knight_generalized_2020} incorporates weighted edges and $k$-hop neighbourhoods. In GNAR models, the value at each node depends not only on its own history but also on a weighted sum of past values from neighbouring nodes up to a fixed network hop, allowing for more flexible spillover effects across nodes without substantially increasing model complexity.

Recent extensions have adopted modelling approaches similar to the GNAR framework for volatility forecasting. \cite{zhang_graph-based_2025} introduced the Graph-HAR (GHAR) model, which augments HAR-style lag structures with neighbourhood-aggregated volatility and correlation measures. In their HAR-DRD framework, the realised covariance matrix is decomposed as $\Sigma_t = D_{t} R_t D_{t}$, where $D_t$ is a diagonal matrix of realised standard deviations and $R_t$ is the realised correlation matrix. Each component is then modelled separately using graph-based predictors. \citet{tapia_costa_higher_2025} build on this by proposing a GNAR-HAR model for forecasting realised covariance matrices, capturing higher-order neighbourhood interactions in both volatilities and correlations. In contrast, our focus is solely on forecasting realised variance, without modelling the full covariance structure or using a DRD decomposition.

Another line of work has focused on incorporating exogenous information into network autoregressive models. While \citet{zhu_network_2017} already noted the possibility of including covariates in NAR-type specifications, \citet{nason_quantifying_2022} formalised this within the GNARX model. GNARX extends GNAR by including node-specific exogenous variables, such as COVID-19 stringency indices and death rates, as additional regressors. Applied to forecasting Purchasing Managers’ Indices across countries using trade-determined networks, GNARX was shown to outperform both standard GNAR and traditional VAR models in terms of predictive accuracy (mean squared error).

Building on these strands of research, we introduce the GNAR-HARX model, a new framework that combines: the network-based cross-sectional modelling of GNAR, the long memory structure of HAR and the flexibility to include node-specific exogenous covariates from GNARX.

The GNAR-HARX model in this case is used to forecast daily realised variance across ten international stock indices, capturing long memory effects, network spillovers, and exogenous influences, often achieving a more parsimonious structure than similar univariate models.

Volatility spillovers are modelled via an underlying network structure, either specified a priori or estimated from historical data. The HAR lags accommodate persistence, and node-level exogenous regressors (e.g., implied volatility indices or overnight returns) allow the model to include more up to date or forward-looking predictive information.

The primary contribution of this thesis is the formal definition and empirical investigation of the GNAR-HARX model, which integrates HAR dynamics and exogenous predictors into the GNAR framework. While related components have been studied individually in previous work, this project brings them together in the context of forecasting realised variance. Building on existing literature, we also propose a set of economically motivated exogenous variables and assess their contribution to predictive performance. We assess the model’s performance using daily realised variance data for ten major international equity indices over the period from 2001 through 2020, benchmarking against established alternatives including HAR, HARX, and GNAR-HAR. To support transparency and future work, we also provide a Python implementation for estimation, forecasting, and model comparison.

The remainder of this thesis is structured as follows:

Section~\ref{sec:model} introduces the GNAR-HARX model in detail, including its mathematical formulation, variants, and estimation procedure. Section~\ref{sec:data} presents and summarises the data used. Section~\ref{sec:methods} outlines the model comparison framework, rolling window design, and evaluation metrics. Section~\ref{sec:results} reports empirical findings, evaluating the forecasting performance of GNAR-HARX and the effect of network structure. Section~\ref{sec:discussion} concludes with a discussion of implications, limitations, and directions for future work.

\section{Model}
\label{sec:model}

\subsection{Model Specification}

\subsubsection{Notation and Definitions}

Let $\mathbf{Y}_t = (Y_{1,t},\ldots,Y_{N,t})^\top$ denote an $N$-variate time series observed over a (possibly time-varying) network $\mathcal{G}_t=(\mathcal{V},\mathcal{E}_t)$. The node set is $\mathcal{V}=\{1,\ldots,N\}$, and the edge set $\mathcal{E}_t \subseteq \mathcal{V}\times\mathcal{V}$ encodes the connections between nodes at time $t$. We use the terms \emph{node} and \emph{vertex} interchangeably. As a motivating example, $Y_{i,t}$ may represent the log realised variance of index $i$ on day $t$. 

Throughout this study, we consider only undirected graphs, so that $(i,j) \in \mathcal{E}_t \iff (j, i) \in \mathcal{E}_t$. The corresponding adjacency matrix $A_t$ has entries $(A_t)_{ij}=1$ if $(i,j) \in \mathcal{E}_t$ and $0$ otherwise. 

For a subset $A \subseteq \mathcal{V}$, the (stage-1) neighbour set of $A$ is defined as
\[
\mathcal{N}_t(A)=\{\,j\in \mathcal{V}\setminus A:\ \exists\, i\in A \text{ with } (i,j)\in\mathcal{E}_t\,\}.
\]

For a single node $i$, write $\mathcal{N}_t^{(1)}(i):=\mathcal{N}_t(\{i\})$. Higher-stage neighbours are defined recursively as
\[
\mathcal{N}_t^{(r)}(i)
=\mathcal{N}_t\!\big(\mathcal{N}_t^{(r-1)}(i)\big)\ \setminus\ \Big(\{i\}\cup \bigcup_{q=1}^{r-1}\mathcal{N}_t^{(q)}(i)\Big),\quad r\ge2.
\]

Alternatively, nodes $i$ and $j$ are $r$-stage neighbours if the shortest path between them in $\mathcal{E}_t$ has distance $r$, i.e.,
$\text{d}_t(i,j)=r$ \citep{nason_modelling_2024}.

For $r\ge 1$ and $j\in \mathcal{N}_t^{(r)}(i)$, let $w^{(r)}_{i,j}(t)\in[0,1]$ denote the connection weight from node $i$ to node $j$ at stage $r$. In this study we adopt uniform (equally weighted) neighbourhood averages:
\[
w^{(r)}_{i,j}(t)=
\begin{cases}
|\mathcal{N}_t^{(r)}(i)|^{-1}, & \text{if } j\in \mathcal{N}_t^{(r)}(i)\ \text{ and }\ |\mathcal{N}_t^{(r)}(i)|>0,\\[3pt]
0, & \text{otherwise}.
\end{cases}
\]

where $|\cdot|$  denotes set cardinality.
Thus $\sum_{j\in \mathcal{N}_t^{(r)}(i)} w^{(r)}_{i,j}(t)=1$ when $\mathcal{N}_t^{(r)}(i)\neq\varnothing$. The adjacency $\mathcal{E}_t$ is specified externally (see Section~\ref{subsec:network_construction} for details on network construction) so that given $\mathcal{E}_t$, the weights are deterministic and not estimated parameters.

For integers $a\ge b\ge 1$, define the non-overlapping lag average
\[
Y_{i,t-a:t-b}\ :=\ \frac{1}{a-b+1}\sum_{k=b}^{a} Y_{i,t-k}.
\]

which is used to form weekly and monthly HAR components.

Let $\{X_{h,i,t}\}$ be the $h$-th stationary exogenous regressor for node $i$ at time $t$, with $h=1,\ldots,H$ and maximum lag $p_h' \in \mathbb{N}_0$.

\subsubsection{GNAR-HAR Model}
The GNAR-HAR model of \cite{tapia_costa_higher_2025} extends the GNAR framework to incorporate Heterogeneous Autoregressive (HAR) dynamics. The GNAR-HAR model describes the evolution of $Y_{i,t}$ for each node $i \in \{1, \dots, N\}$ at time $t \in \{1, \dots, T\}$ as:

\begin{align}
Y_{i,t} &= \alpha_d Y_{i,t-1} + \alpha_w Y_{i,t-5:t-2} + \alpha_m Y_{i,t-22:t-6} \nonumber \\
&\quad + \sum_{r=1}^{r_d} \beta_{d,r} \sum_{j \in \mathcal{N}^{(r)}_t(i)} w^{(r)}_{i,j}(t) Y_{j,t-1}
\nonumber \\
&\quad + \sum_{r=1}^{r_w} \beta_{w,r} \sum_{j \in \mathcal{N}^{(r)}_t(i)} w^{(r)}_{i,j}(t) Y_{j,t-5:t-2} \nonumber \\
&\quad + \sum_{r=1}^{r_m} \beta_{m,r} \sum_{j \in \mathcal{N}^{(r)}_t(i)} w^{(r)}_{i,j}(t) Y_{j,t-22:t-6} + u_{i,t}
\label{eq:gnarhar}
\end{align}

where the error terms $\{u_{i,t}\}$ are mean-zero random variables with variance $\sigma_i^2$.

The autoregressive part follows the HAR structure of \citet{corsi_simple_2009}, using daily, weekly, and monthly non-overlapping averages to capture distinct time scales. The network terms capture spillovers from neighbours at different graph distances.

\subsubsection{GNAR-HARX Model}

We propose the GNAR-HARX model, which extends the GNAR-HAR model in Equation~\ref{eq:gnarhar} by incorporating exogenous regressors. Specifically, the GNAR-HARX model augments Equation~\ref{eq:gnarhar} with the additional term
\[
\sum_{h=1}^{H} \sum_{j'=0}^{p'_h} \lambda_{h,j'} X_{h,i,t-j'}
\]

so that
\begin{align}
Y_{i,t} &= \alpha_d Y_{i,t-1} + \alpha_w Y_{i,t-5:t-2} + \alpha_m Y_{i,t-22:t-6} \nonumber \\
&\quad + \sum_{r=1}^{r_d} \beta_{d,r} \sum_{j \in \mathcal{N}^{(r)}_t(i)} w^{(r)}_{i,j}(t) Y_{j,t-1}
\nonumber \\
&\quad + \sum_{r=1}^{r_w} \beta_{w,r} \sum_{j \in \mathcal{N}^{(r)}_t(i)} w^{(r)}_{i,j}(t) Y_{j,t-5:t-2} \nonumber \\
&\quad + \sum_{r=1}^{r_m} \beta_{m,r} \sum_{j \in \mathcal{N}^{(r)}_t(i)} w^{(r)}_{i,j}(t) Y_{j,t-22:t-6} \nonumber \\
&\quad + \sum_{h=1}^{H} \sum_{j'=0}^{p'_h} \lambda_{h,j'} X_{h,i,t-j'} + u_{i,t}.
\end{align}

Here, the additional term captures the effects of lagged exogenous covariates $X_{h,i,t}$, with $\lambda_{h,j'}$ denoting the effect of the $j'$-th lag of covariate $h$.

\textbf{Parameters:}
\begin{itemize}
    \item $\alpha_d,\alpha_w,\alpha_m$: autoregressive coefficients at daily, weekly, and monthly horizons.
    \item $\beta_{d,r},\beta_{w,r},\beta_{m,r}$: network autoregressive coefficients for neighbourhood stage $r$ at daily, weekly, and monthly horizons.
    \item $\lambda_{h,j'}$: coefficients for lag $j'$ of the $h$-th exogenous variable.
    \item $\mathbf{r}=[r_d,r_w,r_m]$: maximum neighbourhood stages for each time scale.
    \item $\mathcal{N}^{(r)}_t(i)$: stage-$r$ neighbours of node $i$.
    \item $w^{(r)}_{i,j}(t)$: uniform weights across neighbours at stage $r$.

\end{itemize}

\subsubsection{Model Variations and Parameter Counts}
We consider three GNAR-HARX model variants that differ in their degree of parameter sharing across nodes:
\begin{itemize}
    \item Global-$\alpha$ (global): All nodes share autoregressive, network and exogenous parameters.

    \item Local-$\alpha$ (standard): Autoregressive parameters vary by node; network and exogenous parameters are shared.

    \item Local-$\alpha\beta$ (local): All autoregressive, network and exogenous parameters are node-specific.
\end{itemize}

A key advantage of the GNAR-HARX framework lies in its parsimony on large networks, particularly in the global-$\alpha$ and local-$\alpha$ variants. Under the global variant, the autoregressive and network components contribute only $3 + r_d + r_w + r_m$ parameters (three HAR lags plus daily/weekly/monthly network stages), regardless of $N$. Exogenous parameters scale with the number of variables and their lags but, in the global or standard cases, are not multiplied by $N$.

Table~\ref{tab:param_count} summarises the number of parameters under each specification, assuming three HAR lags; daily/weekly/monthly network stages ($r_d$, $r_w$, $r_m$); and $H$ exogenous variables with lag orders $p'_1,\dots,p'_H$:

\begin{table}[H]
\centering
\resizebox{0.9\textwidth}{!}{%
\begin{tabular}{l c c c}
\toprule
\textbf{Model} & \textbf{Autoregressive Terms} & \textbf{Network Terms} & \textbf{Exogenous Terms} \\
\midrule
Global-$\alpha$ & $3$ & $r_d + r_w + r_m$ & $\sum_{h=1}^{H} p'_h $  \\
Local-$\alpha$  & $3N$ & $r_d + r_w + r_m$ & $\sum_{h=1}^{H} p'_h $  \\
Local-$\alpha\beta$ & $3N$ & $N(r_d + r_w + r_m)$ & $N \left(\sum_{h=1}^{H} p'_h \right) $  \\
\bottomrule
\end{tabular}
}
\caption{Parameter counts for GNAR-HARX model variants.}
\label{tab:param_count}
\end{table}

By comparison, fitting $N$ independent HARX models requires $N\!\left(3 + \sum_{h=1}^{H} p'_h\right)$ parameters, which scales linearly with $N$. In contrast, the global-$\alpha$ GNAR-HARX keeps the total parameter count independent of $N$. The local-$\alpha$ specification increases only the autoregressive block to $3N$, while still sharing network and exogenous components across nodes. The more flexible local-$\alpha\beta$ variant scales the autoregressive, network and exogenous parameters with $N$, leading to  higher complexity. Thus, the global and local-$\alpha$ specifications are markedly more parsimonious, which may help to reduce overfitting risk as $N$ grows.

\subsubsection{Stationarity Condition:}

Assuming all exogenous regressors ${X_{h,i,t}}$ are stationary, the GNAR-HARX process $\mathbf{Y}_t$ is stationary if:
\[
\left| \alpha_d \right| + \left| \alpha_w \right| + \left| \alpha_m \right| + \sum_{r=1}^{r_d} \left| \beta_{d,r} \right| + \sum_{r=1}^{r_w} \left| \beta_{w,r} \right| + \sum_{r=1}^{r_m} \left| \beta_{m,r} \right| < 1
\]

A derivation based on the mapping of the GNAR-HARX model to a constrained GNARX process is provided in Appendix~\ref{app:stationarity}.

\section{Data}
\label{sec:data}
This study uses a multivariate time series of daily realised variance and exogenous predictors for ten international equity indices, spanning from \textbf{2 February 2001} to \textbf{30 December 2020}. This sample period is selected because it represents the time frame during which both realised variance and implied volatility data are consistently available for the chosen indices.

\subsection{Data Coverage and Indices}

Table~\ref{tab:vol_indices} lists the ten selected equity indices and their corresponding implied volatility indices. The selection covers major developed markets and reflects a broad geographic and economic representation.

\begin{table}[H]
\centering
\resizebox{\textwidth}{!}{%
\begin{tabular}{llll}
\toprule
\textbf{Stock Index} & \textbf{Index Abbreviation} & \textbf{Volatility Index} & \textbf{Country/Region} \\
\midrule
AEX Index & AEX & VAEX & Netherlands \\
CAC 40 & CAC & VCAC & France \\
DAX & DAX & VDAX-NEW & Germany \\
Dow Jones Industrial Average & DJI & VXD & United States \\
EURO STOXX 50 & STX & VSTOXX & Eurozone \\
FTSE 100 & FTS & VFTSE & United Kingdom \\
Nasdaq 100 & NDX & VXN & United States \\
Nikkei 225 & NKY & VXJ & Japan \\
S\&P 500 Index & SPX & VIX & United States \\
Swiss Market Index & SMI & VSMI & Switzerland \\
\bottomrule
\end{tabular}}
\caption{Stock indices and their associated implied volatility indices}
\label{tab:vol_indices}
\end{table}

Having established the dataset and market coverage, we next describe how the realised variance is measured and constructed.

\subsection{Realised Variance Measures}
\label{subsec: realised_variance_measures}

The realised variance ($RV$) data are sourced from the Oxford-Man Institute’s (OMI) Realized Library \citep{heber_omis_2009}. Specifically, we used the \texttt{rv5\_ss} series, a five-minute subsampled realised variance measure. This estimator, introduced by \citet{zhang_tale_2005}, balances efficiency and robustness by averaging across multiple offset sampling grids to reduce microstructure noise. For clarity, we first outline the theoretical definition of realised variance and then describe how the \texttt{rv5\_ss} series is constructed.

The estimates of $RV$ in the OMI Realized Library implement the methodology developed by \cite{andersen_towards_1998} and formalised in  \cite{andersen_distribution_2001}. These estimates are constructed from high-frequency intraday returns sampled at five-minute intervals, and reflect the ex-post variation in asset prices observed within each trading day.

Formally, let the log-price process of an asset $P_t$ follow:
\[
    dP_t = \mu_t \, dt + \sigma_t \, dW_t,
\]
where $\mu_t$ is a locally bounded drift, $\sigma_t$ is a stochastic volatility process, and $W_t$ is a standard Brownian motion. Over a trading day $t$ of length $H$, with intraday grid
$
0=\tau_{t,0}<\tau_{t,1}<\cdots<\tau_{t,M_t}=H,
$

define the high-frequency log returns on day $t$ by
\[
r_{t,i}:=P_{\tau_{t,i}}-P_{\tau_{t,i-1}},\qquad i=1,\dots,M_t.
\]

\newcommand{\intvar}{\operatorname{IntVar}}
The integrated variance on day $t$ is
\[
    \intvar_t = \int_0^H \sigma_s^2 \, ds
\]
and the realised variance estimator is 
\[
RV_t:=\sum_{i=1}^{M_t} r_{t,i}^2 .
\]

Under infill asymptotics (as the mesh of the grid goes to zero, i.e. $\max_i(\tau_{t,i}-\tau_{t,i-1})\to 0$ with $M_t\to\infty$) and in the absence of microstructure noise or jumps, $RV_t \xrightarrow{p} \intvar_t$ \citep{barndorff-nielsen_econometric_2002}. In practice, we use a finite grid (e.g., five-minute sampling, $M_t \approx 78$ over a 6.5-hour session), so $RV_t$ is only a finite-sample estimator of $\intvar_t$. Compared to squared daily returns, $RV_t$ provides a more efficient and granular measure of return variation by exploiting intraday data and typically improves volatility measurement and forecasting \citep{barndorff-nielsen_econometric_2002,andersen_modeling_2003}.

In practice, however, $RV_t$ is affected by microstructure frictions such as bid–ask bounce and asynchronous trading, which induce noise and bias when sampling at very high frequencies. To mitigate this, \citet{zhang_tale_2005} proposed a subsampling approach that averages realised variance estimates across multiple offset sampling grids, thereby reducing sensitivity to the exact choice of sampling times.

Accordingly, this study adopts the \texttt{rv5\_ss} series from the OMI Realized Library, where the suffix ``ss” denotes subsampling. This measure computes five-minute realised variance across multiple staggered grids and then averages them (Equation~\ref{eqn: rv5_ss}), yielding a daily volatility proxy that is less affected by microstructure noise while retaining efficiency. Formally,

\begin{equation}
\label{eqn: rv5_ss}
    RV_t^{(ss)}=\frac{1}{L} \sum_{\ell=1}^L \sum_{j=1}^M r_{t, j}^{(\ell)2},
\end{equation}

where $r_{t, j}^{(\ell)}$ is the $j$-th five-minute return on grid $\ell$ and $L$ is the number of such grids. 

\begin{figure}[H]
    \centering
    \includegraphics[width=\textwidth]{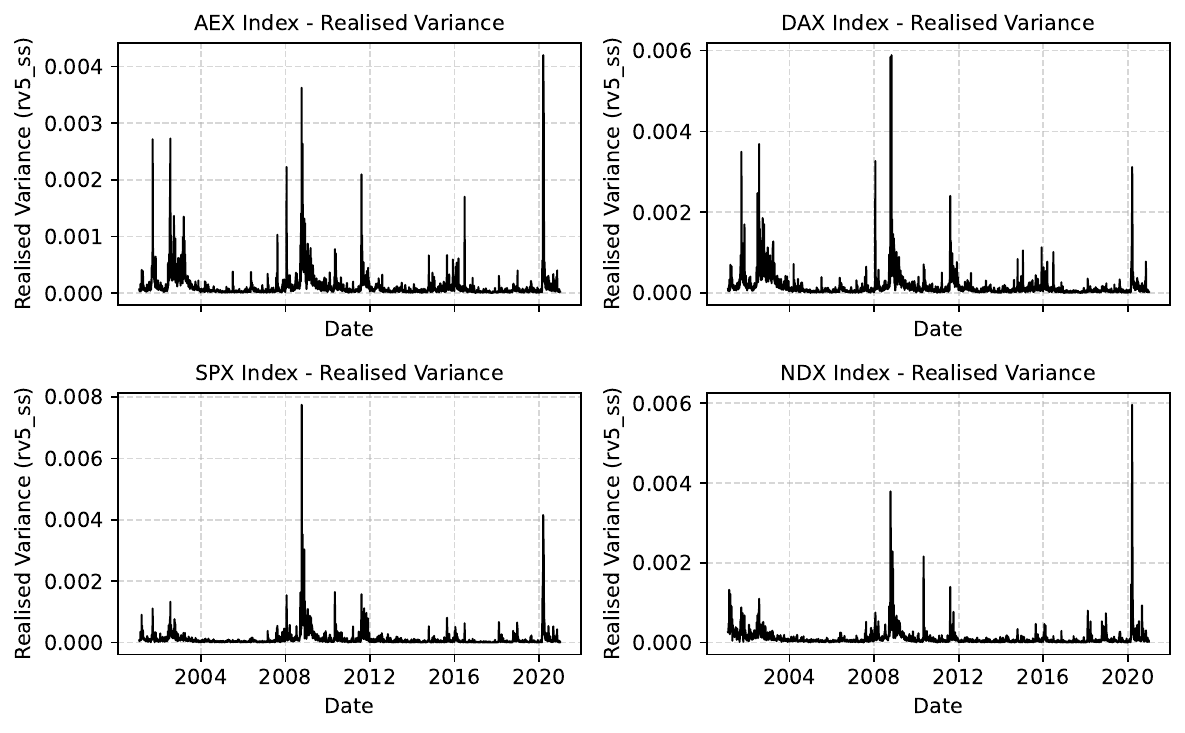}
    \caption{Time series of realised variance ($RV$, measured using the \texttt{rv5\_ss} estimator) for four representative equity indices.}
    \label{fig:rv_panel}
\end{figure}

Figure~\ref{fig:rv_panel} illustrates the \texttt{rv5\_ss} series for selected indices in our dataset (see Table~\ref{tab:vol_indices} for index abbreviations), highlighting the major volatility spikes during the 2008–2009 financial crisis and the 2020 COVID-19 shock.

All subsequent analysis in this study, and references to 
$RV$, are based on the \texttt{rv5\_ss} series.

In addition to realised variance, we also use daily log returns, sourced from the OMI Realized Library, to construct the sparse network via the graphical lasso (see Section~\ref{subsec:network_construction}).

\subsection{Exogenous Variables}

The primary methodological contribution of this study is to augment the GNAR-HAR model with several exogenous variables, motivated by prior literature on volatility forecasting, in an effort to improve forecast accuracy. This study adopts a simplified approach by including only the most recent daily (i.e., one-lag) value of each exogenous variable. This allows us to investigate whether the most up-to-date market information contains incremental predictive power beyond the autoregressive and network components. While previous work often incorporates exogenous variables using HAR-type lag structures (e.g., HAR-IV), exploring such formulations is left for future research.

\begin{itemize}

    \item \textbf{Implied Volatility (IV):} For each stock index, we use its corresponding implied volatility index from Bloomberg (tickers listed in Table~\ref{tab:vol_indices}; e.g., S\&P 500: \texttt{VIX Index}, field \texttt{PX\_LAST}). This represents the market's forward-looking expectation of future volatility over a fixed horizon (typically 30 days). These are computed using option prices and reflect risk-neutral expectations under a model-free framework. The model-free construction of the VIX is based on the theoretical results of \citet{demeter_more_1999}, who show that risk-neutral expected variance can be obtained directly from a range of option prices without assuming a particular asset price model. For example, the VIX Index for the S\&P 500 Index is computed using a wide range of out-of-the-money options.\footnote{See the CBOE VIX White Paper for details: \url{https://cdn.cboe.com/api/global/us_indices/governance/Cboe_Volatility_Index_Mathematics_Methodology.pdf}} Following \cite{busch_role_2011}, we include lagged implied volatility to assess its incremental forecasting power.

    \item \textbf{Asymmetric Returns (Good/Bad Returns):} Motivated by the leverage effect \citep{black_studies_1976, christie_stochastic_1982}, and to allow for asymmetric volatility responses, we decompose daily returns for the stock indices into positive and negative parts:  
    \begin{equation}
        r_t^+ = \max(r_t, 0), \quad r_t^- = \min(r_t, 0),
    \end{equation}
    and include each as a separate regressor. This follows the intuition behind the leverage HAR model of \citet{corsi_discrete-time_2012}, but without using longer-horizon averages.

    \item \textbf{Overnight Returns (ON):} 
    We include lagged overnight returns,
    \begin{equation}
        r_t^{\text{overnight}} = \frac{\text{Open}_t}{\text{Close}_{t-1}} - 1,
    \end{equation}
    to capture information arriving outside regular trading hours. Overnight movements can reflect global macroeconomic news, earnings announcements, or geopolitical shocks.
\citet{wang_volatility_2015} show in Chinese equity markets that negative overnight returns improve volatility forecasts. Consistent evidence is found by \citet{kambouroudis_forecasting_2021}, who report that using overnight returns significantly improves volatility forecasting accuracy for most major international indices.

\end{itemize}

\subsection{Summary Statistics}

Table~\ref{tab:summary_stats} presents summary statistics for the log of the realised variance series (as defined in Equation~\ref{eqn: rv5_ss}) of each index. The series exhibit strong persistence, as indicated by the high first-order autocorrelations (ranging from 0.77 to 0.86). The distributions also show moderate positive skewness, and all exhibit leptokurtosis (kurtosis ranging from 3.26 to 4.55), reflecting fat tails relative to the normal distribution, a well-documented feature of financial volatility. The SMI index stands out with the highest skewness (1.07) and kurtosis (4.55), suggesting occasional large spikes in volatility. These stylised facts support the use of long-memory or HAR-type models.

\begin{table}[ht]
\centering
\resizebox{\textwidth}{!}{%
\begin{tabular}{lcccccccccc}
\toprule
\textbf{Index} & \textbf{Mean} & \textbf{Std. Dev.} & \textbf{Skew.} & \textbf{Kurt.} & \textbf{ACF(1)} & \textbf{ACF(2)} & \textbf{ACF(3)} & \textbf{PACF(1)} & \textbf{PACF(2)} & \textbf{PACF(3)} \\
\midrule
AEX & -9.644 & 1.015 & 0.639 & 3.376 & 0.856 & 0.819 & 0.797 & 0.857 & 0.320 & 0.193 \\
CAC & -9.446 & 0.977 & 0.522 & 3.317 & 0.841 & 0.803 & 0.782 & 0.841 & 0.328 & 0.200 \\
DAX & -9.348 & 1.031 & 0.520 & 3.319 & 0.845 & 0.808 & 0.789 & 0.845 & 0.329 & 0.211 \\
DJI & -9.862 & 1.129 & 0.521 & 3.563 & 0.795 & 0.763 & 0.730 & 0.795 & 0.356 & 0.175 \\
STX & -9.317 & 1.004 & 0.533 & 3.492 & 0.818 & 0.774 & 0.753 & 0.818 & 0.318 & 0.204 \\
FTS & -9.647 & 1.010 & 0.693 & 3.712 & 0.785 & 0.751 & 0.728 & 0.785 & 0.353 & 0.207 \\
NDX & -9.741 & 1.047 & 0.476 & 3.261 & 0.833 & 0.780 & 0.745 & 0.833 & 0.282 & 0.155 \\
NKY & -9.702 & 0.950 & 0.410 & 3.521 & 0.773 & 0.722 & 0.696 & 0.773 & 0.309 & 0.196 \\
SPX & -9.894 & 1.154 & 0.476 & 3.482 & 0.823 & 0.783 & 0.748 & 0.823 & 0.328 & 0.159 \\
SMI & -9.890 & 0.912 & 1.068 & 4.554 & 0.856 & 0.821 & 0.803 & 0.856 & 0.331 & 0.211 \\
\bottomrule
\end{tabular}
}
\caption{Summary statistics of $\log RV$ for each index}
\label{tab:summary_stats}
\end{table}


\section{Methods}
\label{sec:methods}

This section describes the forecasting framework, estimation procedure, network construction, and evaluation metrics used to assess GNAR-HAR(X) models for predicting the daily realised variance ($RV$) of ten major international stock indices. We benchmark these multivariate models against univariate HAR and HARX specifications, evaluated for both predictive accuracy and parameter efficiency.
The HAR and HARX models are estimated using the \texttt{HARX} implementation from the \texttt{arch} Python package. This study provides the first direct comparison of GNAR-HAR(X) models with HAR/HARX models in this setting.

\subsection{Forecasting Framework}
\label{subsec:framework}
Let $RV_{i, t}$ denote the realised variance of asset $i$ on day $t$, constructed as described in Section~\ref{subsec: realised_variance_measures}. We model the log-transformed realised variance, $Y_{i,t} = \log RV_{i,t}$. Empirical results in \citet{clements_practical_2021} suggest that log-transformation improves forecast accuracy over using raw or square-root-transformed variance.

Each model is trained on an initial in-sample window of four years to avoid look-ahead bias (i.e., the use of future information that would not have been available at the time forecasts are made), which can lead to overly optimistic performance estimates. 

For GNAR-HAR(X) models using graphical lasso (GL) networks, the initial four-year in-sample window is also used to estimate the first network structure (see Section~\ref{subsec:network_construction} for details). At each subsequent refit step in the rolling window procedure, the GL network is re-estimated on the updated three-year window, allowing the network to evolve with the data.

We implement a rolling window forecasting procedure, repeating the following steps until the end of the evaluation period:

\begin{itemize}
\item Refit the model using the most recent three years of data (approximately 756 trading days). For models using GL networks, the network is re-estimated on this updated window.
\item Generate one-step-ahead forecasts for the next 22 trading days.
\item Shift the window forward by 22 trading days and repeat.
\end{itemize}

At each refit step, the response and all regressors are standardised within the three-year training window by subtracting their sample mean and dividing by their sample standard deviation. This normalisation ensures comparability across predictors and stabilises the estimation procedure.

This rolling procedure produces one-day-ahead forecasts across the out-of-sample evaluation period for each index. Estimation of the models requires only that the innovations ${u_{i,t}}$ are mean-zero with finite variance. However, when transforming the forecasts from $\log RV$ back to $RV$, an additional distributional assumption is needed to correct for Jensen’s inequality, since the exponential function is convex. We assume Gaussian residuals so that under this assumption, the unbiased estimator of $RV$ is (see Appendix~\ref{app:jensen} for details):
\begin{equation}
\widehat{RV}_{i,t} = \exp\left( \widehat{Y}_{i,t} + \tfrac{1}{2}\widehat{\sigma_i}^2 \right),
\label{eq:jensen_adjustment}
\end{equation}
where $\widehat{\sigma_i}^2$ is the in-sample residual variance from the rolling window.

In all GNAR-HAR(X) models, we restrict attention to first-order neighbourhoods, setting $r_d = r_w = r_m = 1$ for daily, weekly, and monthly spillovers. This approach, which follows \cite{zhang_graph-based_2025}, is motivated by two considerations. First, in fully connected networks all nodes are directly linked, so higher-order neighbourhoods do not expand the information set. To ensure differences between fully connected and GL-based models reflect only the network structure, rather than neighbourhood orders, we also enforce $r=1$ in the GL case. Second, because a large number of model variants and combinations of exogenous predictors are already under comparison, we fix $r=1$ to avoid expanding the model search space further. Preliminary experiments with $r > 1$ showed no consistent gains in forecast accuracy, similar to the findings of \cite{zhang2025forecasting}, supporting this restriction. While alternative selection strategies, such as information criteria (e.g., BIC) or graphical diagnostics like the Corbit plot of \citet{nason_new_2023}, could be considered, we adopt a fixed $r=1$ specification.

\subsection{Network Construction}
\label{subsec:network_construction}
We consider two network structures for the GNAR-based models.

\textbf{(i) Fully Connected Network}

Each node is connected to every other node (referred to as a complete graph in the graph theory literature). This choice reflects the high degree of integration in global financial markets, an effect observed most strongly during periods of market stress \citep{korkusuz_complex_2023}.

\textbf{(ii) Sparse Network via Graphical Lasso}

To obtain a sparser, data-driven network, we estimate the network from the inverse covariance (precision) matrix of daily log returns by applying the graphical lasso \citep{friedman_sparse_2008}. The graphical lasso solves the optimisation problem:
\begin{equation}
\hat{\Theta} = \arg\max_{\Theta \succ 0} \left\{ \log\det(\Theta) - \mathrm{tr}(S \Theta) - \rho |\Theta|_1 \right\},
\label{eq:glasso}
\end{equation}

where $S$ is the sample covariance matrix of the standardised log returns, $\Theta$ is the precision matrix (inverse of the covariance matrix), and $\rho$ is a regularisation parameter controlling sparsity via the $\ell_1$ norm.

In this study, the graphical lasso is implemented using the \texttt{GraphicalLasso} and \\ \texttt{GraphicalLassoCV} classes from the \texttt{scikit-learn} Python library.

To avoid look-ahead bias, $\rho$ is selected using ten-fold cross-validation on the initial four-year in-sample period. This chosen value is then held fixed throughout the forecasting procedure. At each refit, the graphical lasso is re-applied to the most recent three-year window with this fixed $\rho$, allowing the network to adapt over time while keeping the regularisation level consistent.

The adjacency matrix is then obtained by assigning an edge between nodes $i$ and $j$ whenever the corresponding off-diagonal entry of $\hat{\Theta}$ is non-zero, indicating conditional dependence between their daily log return series given all others. An example of a network produced by the graphical lasso is provided in Figure~\ref{fig:gl_network} (Appendix).

\subsection{Evaluation Metrics}
\label{subsec:evaluation_metrics}
We assess model performance using two standard loss functions, computed on forecast errors $e_{i,t}=RV_{i,t}- \widehat{RV}_{i,t} $ and averaged over all indices $i=1,\dots,N$ and dates $t=1,\dots,T$:

\begin{enumerate}
    \item Mean Squared Error (MSE):
    \begin{equation}
        \text{MSE} = \frac{1}{NT} \sum_{i=1}^{N} \sum_{t=1}^{T} e_{i,t}^{2}.
        \label{eq:mse}
    \end{equation}

    \item Quasi-Likelihood (QLIKE):
    \begin{equation}
        \text{QLIKE} = \frac{1}{NT} \sum_{i=1}^{N} \sum_{t=1}^{T}
        \left[ \log \widehat{RV}_{i,t} + \frac{RV_{i,t}}{\widehat{RV}_{i,t}} \right].
        \label{eq:qlike}
    \end{equation}
\end{enumerate}

QLIKE is robust to measurement error in volatility and penalises large forecast errors less severely than MSE, particularly in the right tail. However, it is asymmetric, tending to favour positively biased forecasts over equally sized negative ones \citep{patton_volatility_2011}. While QLIKE is often preferred in financial volatility contexts, reporting both metrics offers a more balanced assessment of forecasting performance.

\section{Results}
\label{sec:results}

This section presents the empirical results from our forecasting study using the GNAR-HARX model and its benchmarks. Models are evaluated on one-step-ahead forecasting performance using mean squared error (MSE) and QLIKE loss within the rolling window framework described in Section~\ref{subsec:framework}.

\begin{table}[htbp]
\centering
\scriptsize

\begin{tabular}{llllrrr}
\toprule
\textbf{Model} & \textbf{Variant} & \textbf{Network} & \textbf{Exogenous Variables} & \textbf{QLIKE} & \textbf{Rel QLIKE} & \textbf{Rel MSE} \\
\midrule
GNAR-HAR & local & FC & [] & -8.5891 & 1.00 & 1.00 \\
GNAR-HARX & local & FC & [`on'] & -8.5889 & 1.00 & 1.00 \\
GNAR-HAR & standard & FC & [] & -8.5884 & 1.00 & 1.01 \\
GNAR-HARX & standard & FC & [`on'] & -8.5882 & 1.00 & 1.01 \\
GNAR-HARX & standard & FC & [`iv'] & -8.5881 & 1.00 & 1.00 \\
GNAR-HARX & local & FC & [`good', `bad'] & -8.5881 & 1.00 & 1.01 \\
GNAR-HARX & standard & FC & [`iv', `on'] & -8.5880 & 1.00 & 1.00 \\
GNAR-HARX & local & FC & [`good', `bad', `on'] & -8.5877 & 1.00 & 1.01 \\
GNAR-HARX & local & FC & [`iv'] & -8.5876 & 1.00 & 1.01 \\
GNAR-HAR & global & FC & [] & -8.5876 & 1.00 & 1.01 \\
GNAR-HARX & global & FC & [`iv'] & -8.5876 & 1.00 & 1.00 \\
GNAR-HARX & standard & FC & [`good', `bad'] & -8.5875 & 1.00 & 1.01 \\
GNAR-HARX & global & FC & [`on'] & -8.5874 & 1.00 & 1.01 \\
GNAR-HARX & local & FC & [`iv', `on'] & -8.5874 & 1.00 & 1.01 \\
GNAR-HARX & global & FC & [`iv', `on'] & -8.5874 & 1.00 & 1.00 \\
GNAR-HARX & standard & FC & [`iv', `good', `bad'] & -8.5873 & 1.00 & 1.00 \\
GNAR-HARX & standard & FC & [`good', `bad', `on'] & -8.5872 & 1.00 & 1.01 \\
GNAR-HARX & standard & FC & [`iv', `good', `bad', `on'] & -8.5871 & 1.00 & 1.00 \\
GNAR-HARX & local & FC & [`iv', `good', `bad'] & -8.5869 & 1.00 & 1.02 \\
GNAR-HAR & local & GL & [] & -8.5867 & 1.00 & 1.02 \\
GNAR-HARX & global & FC & [`good', `bad'] & -8.5867 & 1.00 & 1.01 \\
GNAR-HARX & global & FC & [`iv', `good', `bad'] & -8.5867 & 1.00 & 1.01 \\
GNAR-HARX & local & FC & [`iv', `good', `bad', `on'] & -8.5866 & 1.00 & 1.02 \\
GNAR-HARX & local & GL & [`on'] & -8.5865 & 1.00 & 1.02 \\
GNAR-HARX & global & FC & [`iv', `good', `bad', `on'] & -8.5864 & 1.00 & 1.01 \\
GNAR-HARX & global & FC & [`good', `bad', `on'] & -8.5864 & 1.00 & 1.02 \\
GNAR-HARX & local & GL & [`good', `bad'] & -8.5857 & 1.00 & 1.02 \\
GNAR-HAR & standard & GL & [] & -8.5856 & 1.00 & 1.03 \\
GNAR-HARX & standard & GL & [`iv'] & -8.5854 & 1.00 & 1.02 \\
GNAR-HARX & standard & GL & [`on'] & -8.5854 & 1.00 & 1.03 \\
GNAR-HARX & local & GL & [`iv'] & -8.5854 & 1.00 & 1.02 \\
GNAR-HARX & local & GL & [`good', `bad', `on'] & -8.5852 & 1.00 & 1.02 \\
GNAR-HARX & standard & GL & [`iv', `on'] & -8.5852 & 1.00 & 1.02 \\
GNAR-HARX & local & GL & [`iv', `on'] & -8.5851 & 1.00 & 1.02 \\
GNAR-HAR & global & GL & [] & -8.5849 & 1.00 & 1.03 \\
GNAR-HARX & global & GL & [`iv'] & -8.5848 & 1.00 & 1.02 \\
GNAR-HARX & global & GL & [`on'] & -8.5846 & 1.00 & 1.03 \\
GNAR-HARX & standard & GL & [`good', `bad'] & -8.5846 & 1.00 & 1.03 \\
GNAR-HARX & global & GL & [`iv', `on'] & -8.5845 & 1.00 & 1.02 \\
GNAR-HARX & local & GL & [`iv', `good', `bad'] & -8.5845 & 1.00 & 1.03 \\
GNAR-HARX & standard & GL & [`iv', `good', `bad'] & -8.5845 & 1.00 & 1.02 \\
GNAR-HARX & standard & GL & [`good', `bad', `on'] & -8.5843 & 1.00 & 1.03 \\
GNAR-HARX & standard & GL & [`iv', `good', `bad', `on'] & -8.5842 & 1.00 & 1.02 \\
GNAR-HARX & local & GL & [`iv', `good', `bad', `on'] & -8.5841 & 1.00 & 1.03 \\
GNAR-HARX & global & GL & [`good', `bad'] & -8.5839 & 1.00 & 1.03 \\
GNAR-HARX & global & GL & [`iv', `good', `bad'] & -8.5838 & 1.00 & 1.02 \\
GNAR-HARX & global & GL & [`iv', `good', `bad', `on'] & -8.5835 & 1.00 & 1.03 \\
GNAR-HARX & global & GL & [`good', `bad', `on'] & -8.5835 & 1.00 & 1.03 \\
HARX & local & None & [`iv'] & -8.5831 & 1.00 & 17139.38 \\
HARX & local & None & [`iv', `good', `bad'] & -8.5824 & 1.00 & 8669.26 \\
HARX & local & None & [`iv', `on'] & -8.5823 & 1.00 & 1763807.26 \\
HARX & local & None & [`iv', `good', `bad', `on'] & -8.5821 & 1.00 & 616700.95 \\
HAR & local & None & [] & -8.5785 & 1.00 & 1.10 \\
HARX & local & None & [`good', `bad'] & -8.5777 & 1.00 & 1.47 \\
HARX & local & None & [`good', `bad', `on'] & -8.5771 & 1.00 & 10.24 \\
HARX & local & None & [`on'] & -8.5759 & 1.00 & 13.31 \\
\bottomrule
\end{tabular}

\caption{Forecasting performance of GNAR-HARX and benchmark models across 
configurations, ordered by QLIKE from best to worst.
Exogenous abbreviations: \texttt{iv} = implied volatility, 
\texttt{good} and \texttt{bad} = positive and negative components of the previous return, 
and \texttt{on} = overnight return. Networks: \texttt{FC} = fully connected, \texttt{GL} = graphical lasso.}
\label{tab:full-results}

\end{table}

\subsection{Model Comparison}
\label{sec:model-comparison}

Table~\ref{tab:full-results} summarises forecasting performance across all models, ordered by QLIKE. Relative performance is expressed as the ratio to the best-performing model for each respective metric, with both QLIKE and MSE reported.

The best overall model as judged by QLIKE is the \textbf{local GNAR-HAR} with a fully connected (FC) network and no exogenous regressors. By MSE, the top performer is the \textbf{standard GNAR-HARX} with implied volatility (IV) as the sole exogenous variable, also with an FC network. Differences from other leading GNAR-HAR(X) specifications are minimal, with relative QLIKE and MSE values almost indistinguishable to two decimal places.

A consistent pattern is that local and standard GNAR-HAR(X) variants slightly outperform the global version, though the gap is very small. The global specification achieves nearly identical performance but with fewer parameters.

Among GNAR-HAR(X) specifications, the top models (as judged by QLIKE) for each variant exclude exogenous predictors, suggesting that once network-aggregated lags are accounted for, external variables add little incremental value. Nevertheless, when exogenous variables are included, overnight returns and implied volatility are more often associated with stronger performance, whereas the leverage effect (captured by good/bad returns) plays a more limited role.

The univariate HAR and HARX benchmarks underperform the GNAR-based models in terms of both QLIKE and MSE. This is despite their requirement of many more parameters in some cases (e.g., a total of 40 across nodes in the HARX model with IV, compared to just 7 in the global GNAR-HARX model with IV). The best HARX specification with IV achieves a QLIKE of $-8.5831$, close to but still below the leading GNAR-HARX models, and shows substantial instability in MSE. In particular, the HARX models using IV as an exogenous variable record extremely large relative MSE values (e.g., exceeding $1{,}000$ or even $1{,}000{,}000$) despite otherwise reasonable QLIKE. This behaviour is concentrated around 17 March 2020, when forecasts incorporated the VIX closing value from the previous day. On Monday, 16 March 2020 the VIX closed at 82.69, its highest closing level on record (at the time of writing), exceeding the previous high of 80.86 set on 20 November 2008 and reflecting acute market stress following the U.S. national emergency declaration on Friday, 13 March \citep{apergis_role_2023}. This value was well above levels observed in the refit window ending 14 February 2020, as shown in Figure~\ref{fig:iv_vs_rv} (Appendix). Because HARX is linear in $\log RV$, such an extreme IV input drives an unusually large shift in the linear predictor, and after the exponential back-transform this leads to outsized forecasts for $RV$. MSE penalises these forecasts quadratically, whereas QLIKE is less sensitive to extreme over-predictions, explaining the divergence in the two metrics.

By contrast, GNAR-HARX models that include IV remain stable in terms of MSE. Their estimated IV coefficients are materially smaller than in the univariate HARX, as shown in Figure~\ref{fig:iv_harx_vs_gnarharx} (Appendix). This suggests that the network-lag ($\beta$) terms already capture much of the common volatility factor that IV proxies. As a result, IV contributes only marginally on top of the network terms, and the forecasts avoid the extreme MSE spikes seen in the univariate case.

It is notable that the top-performing models, as judged by both QLIKE and MSE, use the FC network. In contrast, models using graphical lasso (GL) networks consistently rank lower, even when augmented with the same exogenous variables. These results imply that, in this setting, the potential benefits of data-driven sparsity appear outweighed by estimation challenges, leading to inferior predictive performance.

\subsection{Coefficient Dynamics}
\label{sec:coefficient-dynamics}
\subsubsection{Alpha and Beta Coefficients}
\label{subsubsec:network_coefs}
\begin{figure}[H]
    \centering
    \includegraphics[width=0.8\textwidth]{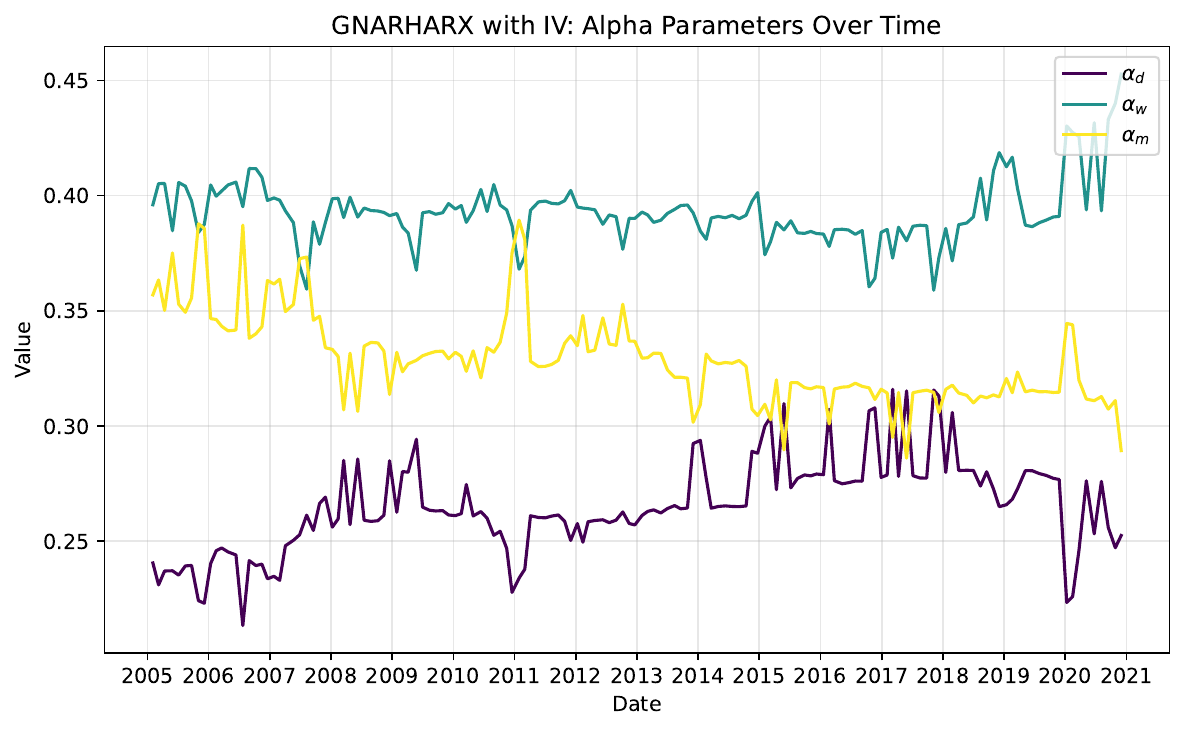}
    \caption{Evolution of $\alpha$ coefficients over time for the best global GNAR-HARX model.}
    \label{fig:alpha-over-time}
\end{figure}

To further understand the behaviour of the best global GNAR-HARX model (FC network with IV), we examine the evolution of its estimated parameters over the full out-of-sample evaluation period (February 2005 - December 2020), using the rolling-window estimates re-fitted at each step. Figure~\ref{fig:alpha-over-time} plots the estimated $\alpha$ parameters, which capture the relative contribution of daily, weekly, and monthly lags of $\log RV$ for the same node. All three coefficients are positive and relatively stable over time, with $\alpha_w$ (weekly) generally dominating. This indicates that mid-range memory effects are the most persistent driver of volatility dynamics for this particular model.

\begin{figure}[ht]
    \centering
    \includegraphics[width=0.8\textwidth]{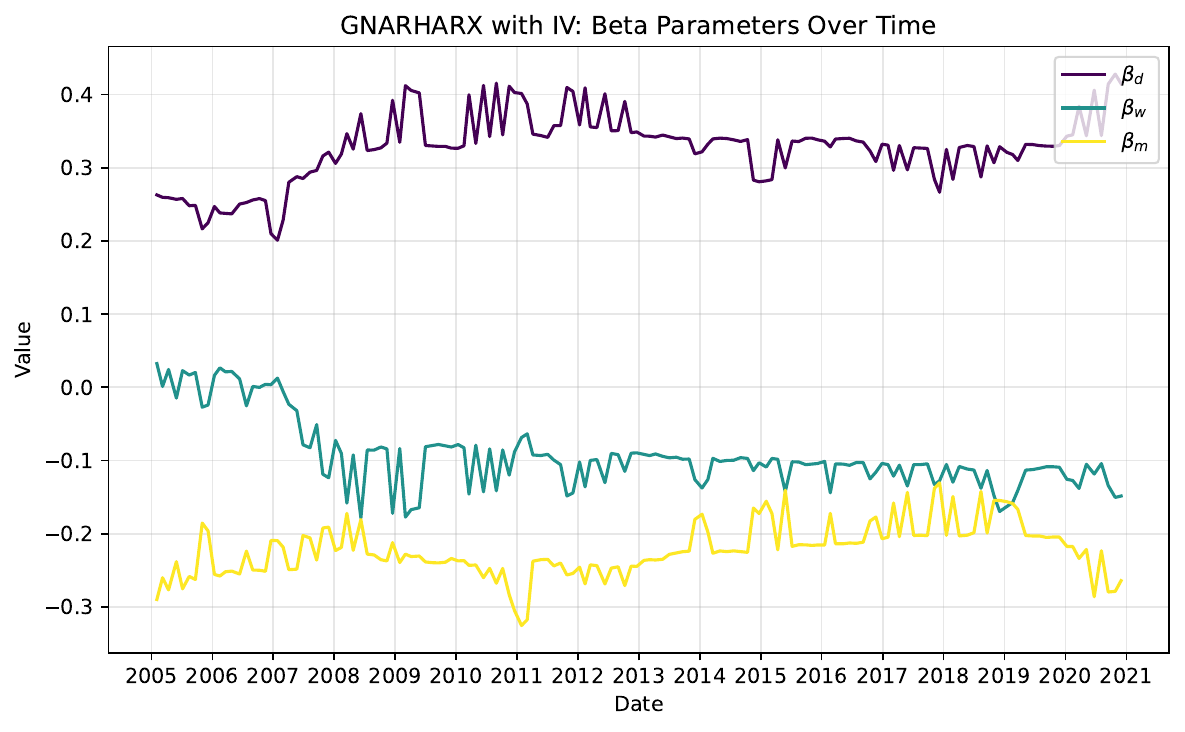}
    \caption{Evolution of $\beta$ coefficients over time for the best global GNAR-HARX model.}
    \label{fig:beta-over-time}
\end{figure}

Figure~\ref{fig:beta-over-time} shows the corresponding $\beta$ parameters, which reflect lagged volatility spillovers from connected nodes. Here we observe more variability and some structural shifts. For example, $\beta_d$ (daily) increases notably around 2007–2010, suggesting a stronger role for contemporaneous spillovers during the global financial crisis, and rises again post-2020 during the COVID-19 pandemic and associated market turbulence. This may indicate that market participants shifted their attention from longer horizons to daily signals, reflecting heightened sensitivity to new information during crises. By contrast, the negative values of $\beta_w$ and $\beta_m$ may point to mean-reverting spillover effects across nodes at these horizons.

\subsubsection{Exogenous Coefficients}
\label{subsubsec:exog_coefs}
\begin{figure}[ht]
    \centering
    \includegraphics[width=0.8\textwidth]{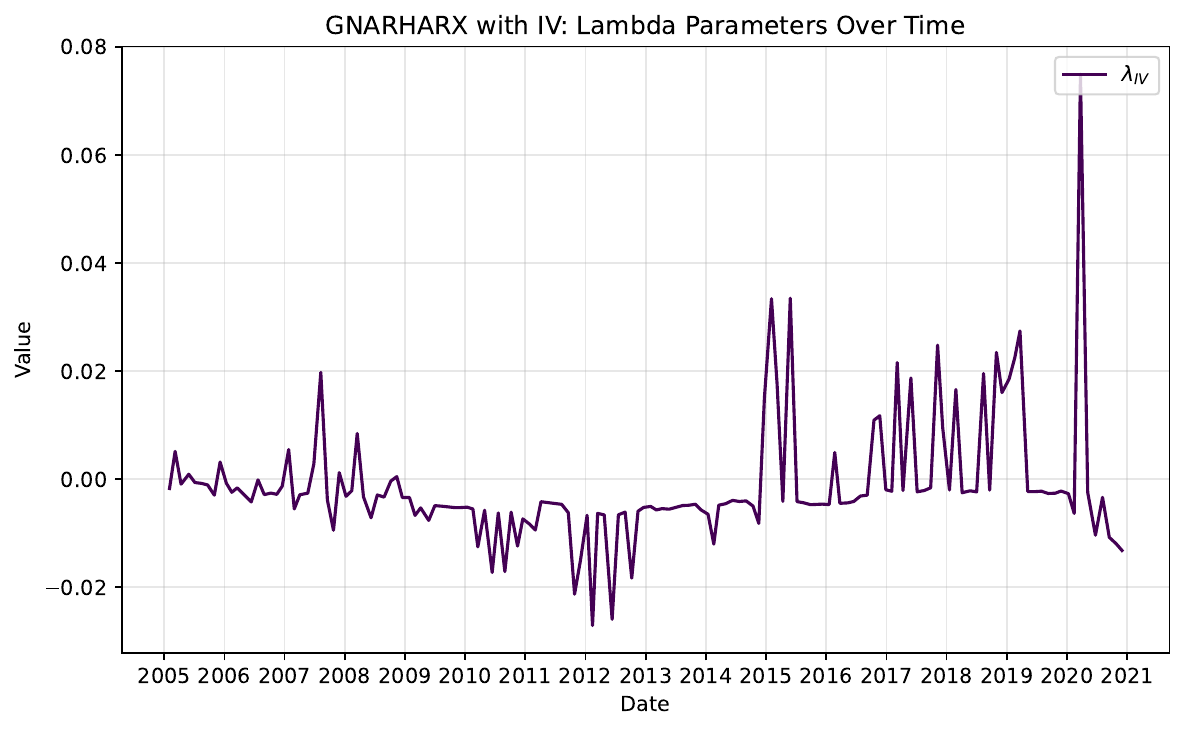}
    \caption{Time-varying exogenous coefficients in the best global GNAR-HARX model using implied volatility.}
    \label{fig:exog-coeffs}
\end{figure}

Finally, Figure~\ref{fig:exog-coeffs} presents the exogenous coefficient $\lambda_{\text{IV}}$ for implied volatility. Although relatively small in magnitude throughout, the coefficient becomes more pronounced on several occasions, most notably when the model is refitted on 23 March 2020, coinciding with the impact of the COVID-19 pandemic on financial markets. This aligns with the notion that IV can carry strong predictive power during periods of market uncertainty.

\subsection{Residual Analysis}
\label{sec:residual-analysis}

To further evaluate the performance of the best global GNAR-HARX model (FC network with IV), we conduct a residual analysis for a representative node, the SPX Index (S\&P 500).

\begin{figure}[ht]
    \centering
    \includegraphics[width=0.8\textwidth]{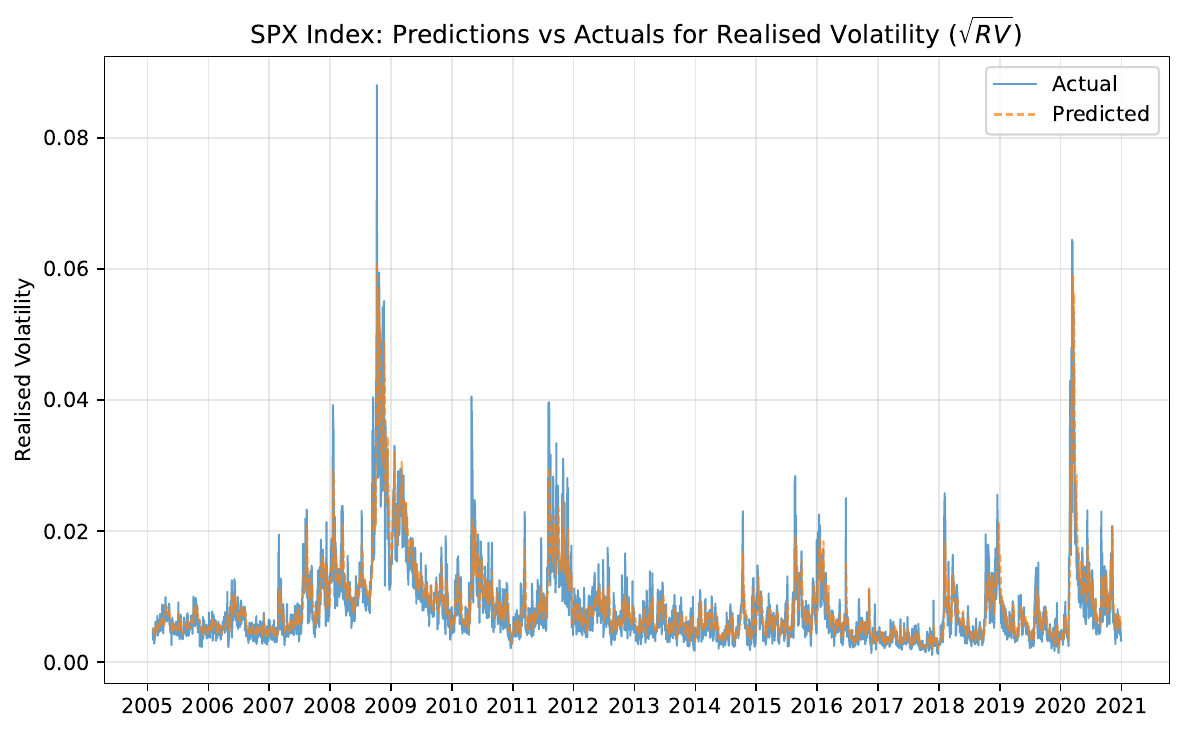}
    \caption{Predicted vs actual realised volatility ($\sqrt{RV}$) for the SPX Index using the best global GNAR-HARX model.}
    \label{fig:pred-vs-actual-spx}
\end{figure}

Figure~\ref{fig:pred-vs-actual-spx} presents predicted versus actual values for realised volatility, i.e. the square-root of realised variance (\(\sqrt{RV}\)). This transformation is commonly used for visualisation because it compresses large spikes while preserving interpretability in volatility units. Visually, the model tracks periods of market turbulence (e.g., the 2008 crisis and 2020 COVID shock) reasonably well, though some underestimation is evident at extreme peaks.

\begin{figure}[H]
    \centering
    \includegraphics[width=0.8\textwidth]{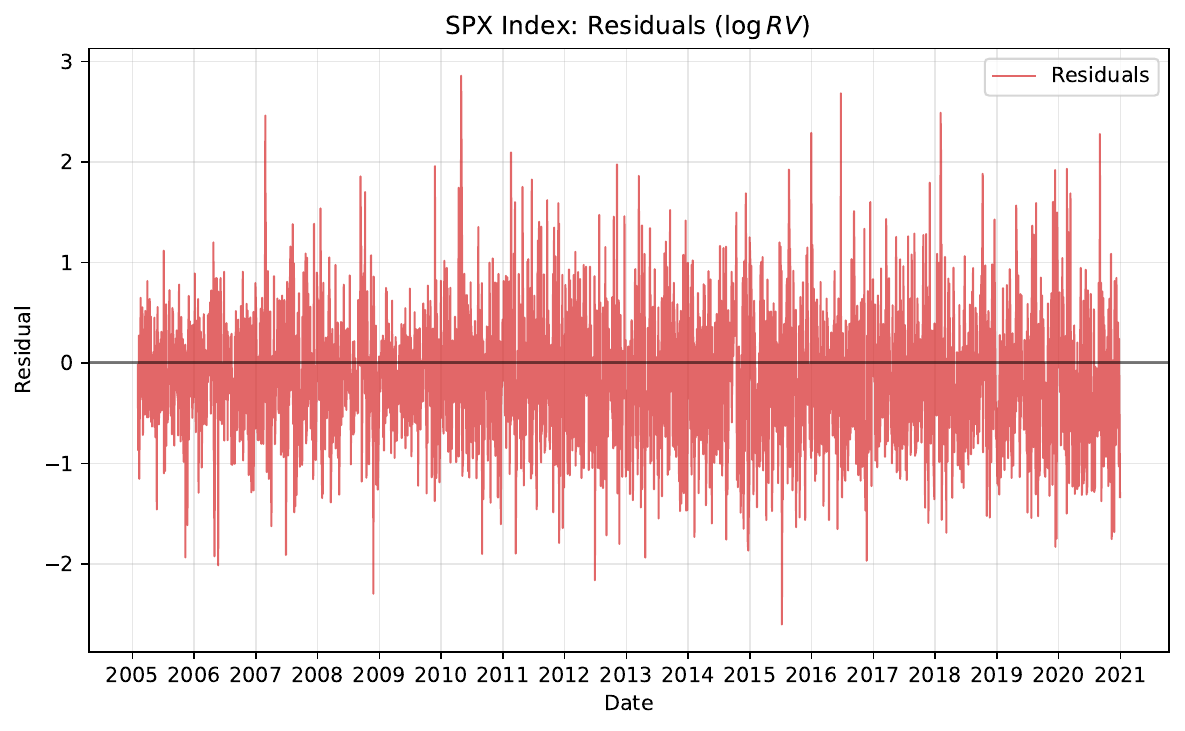}
    \caption{Residuals (log-realised variance) over time for the SPX Index.}
    \label{fig:residuals-time-spx}
\end{figure}

\begin{figure}[ht]
    \centering
    \includegraphics[width=0.7\textwidth]{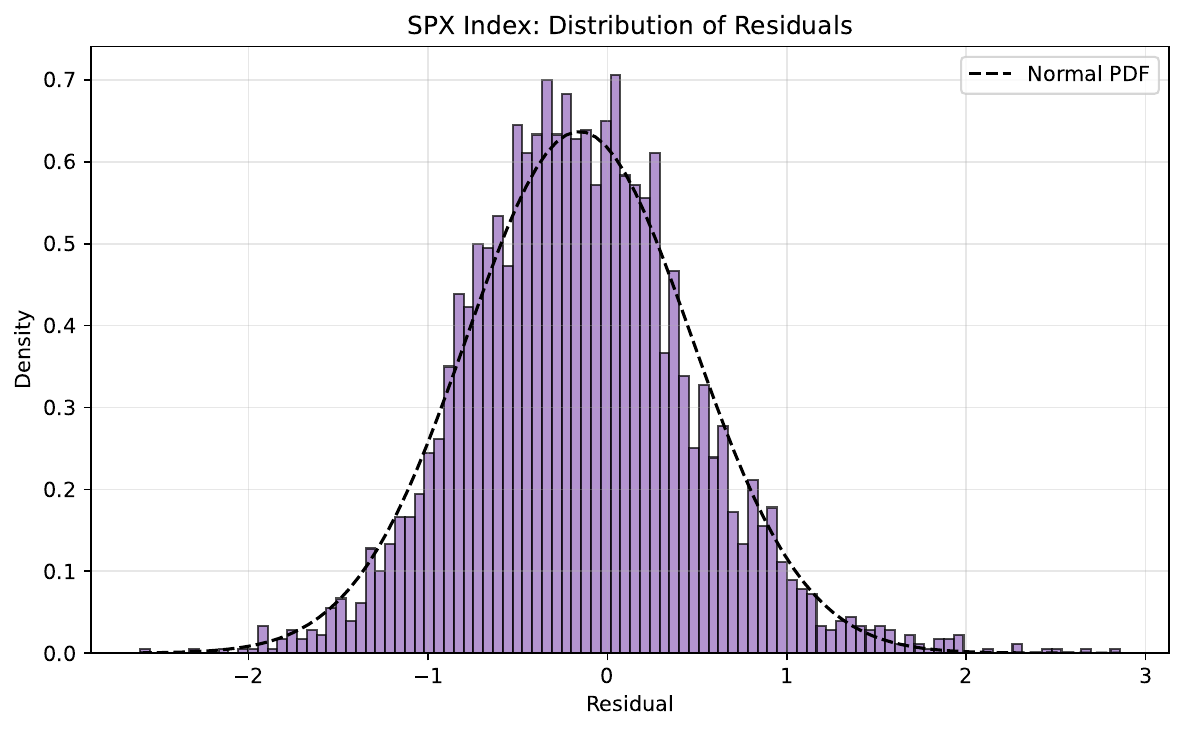}
    \caption{Distribution of residuals (log-realised variance) for the SPX Index, with overlaid normal density.}
    \label{fig:residuals-distribution-spx}
\end{figure}

Figures~\ref{fig:residuals-time-spx} and \ref{fig:residuals-distribution-spx} show residuals in the log-realised variance space, the scale on which the model was estimated. Analysing residuals on this scale is more appropriate, as it aligns with the assumed error distribution and ensures that diagnostic properties (e.g., normality and mean-zero errors) reflect the model’s true fitting behaviour.

The residual time series (Figure~\ref{fig:residuals-time-spx}) is centred around zero, with occasional large residuals during volatile periods. This suggests that while the model may adapt well to regular market conditions, some extreme shocks are not fully captured.

The histogram in Figure~\ref{fig:residuals-distribution-spx} shows a roughly symmetric, bell-shaped distribution with heavier tails than a normal distribution, particularly on the right. The slight positive skew and excess kurtosis suggest occasional under-prediction of realised variance, motivating the consideration of heavier-tailed error distributions.

Similar analyses for the remaining indices are provided in Appendix~\ref{app:residual-other}. These confirm that residuals are generally centred around zero (Figure~\ref{fig:residuals-all}) and approximately bell-shaped with heavier tails than normality would suggest (Figure~\ref{fig:residuals-all-dist}). This indicates that the diagnostic features observed for the SPX Index also hold across the full panel.

\subsection{Network Structure Analysis}
\label{sec:network-analysis}

We further explore the performance gap between models using FC and GL networks. While the GL network is dynamically refitted at each training window to capture time-varying relationships between indices, this flexibility does not translate into improved forecasting performance. This may be due to instability or over-fitting in sparse graphs, especially in high-volatility periods.

\subsection{Graphical Lasso Network Dynamics}
\label{sec:network-dynamics}

To better understand the evolving structure of the GL network, we investigate the evolution of the adjacency matrices used. Although these networks are refitted monthly to capture time-varying relationships between nodes, their use did not consistently improve out-of-sample performance relative to the simpler FC network.

\begin{figure}[ht]
    \centering
    \includegraphics[width=0.8\textwidth]{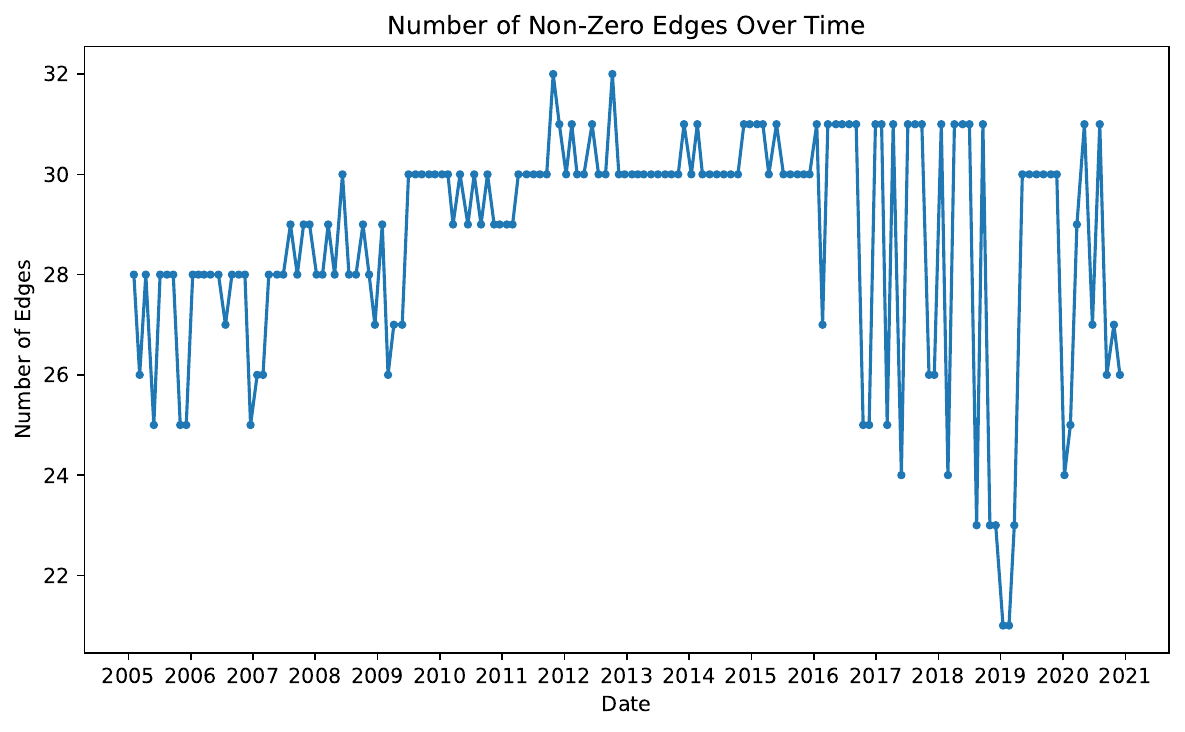}
    \caption{Number of non-zero edges in the GL network over time, reflecting changing estimated market connectivity.}
    \label{fig:gl-edge-count}
\end{figure}

Figure~\ref{fig:gl-edge-count} shows the number of non-zero edges in the graphical lasso network over time. From 2005 through 2015, the number of active connections remains relatively stable, ranging between 26 and 32, suggesting a relatively consistent network structure during this period. However, from 2016 onward, we observe more variance in edge count, with sharp drops in connectivity, especially from 2018 to 2020. These changes may be driven either by over-shrinkage of the penalised estimator in noisy conditions or by genuine structural breaks.

To measure structural persistence, we compute the Jaccard similarity index between consecutive networks. The Jaccard index is well suited to this task because it directly quantifies the proportion of common edges across two networks, relative to the total number of distinct edges. Unlike raw edge counts, it provides a scale-free measure of overlap that lies in $[0,1]$, enabling us to track the degree of structural stability or turnover in the estimated adjacency matrices over time. Specifically, for two binary adjacency matrices $A^{(t)}$ and $A^{(t-1)}$, we first extract the upper-triangular entries (excluding the diagonal) to avoid double-counting. Let $\mathbf{a}^{(t)}$ and $\mathbf{a}^{(t-1)}$ denote these binary edge indicator vectors. The Jaccard similarity is then defined as
\[
J\big(\mathbf{a}^{(t)}, \mathbf{a}^{(t-1)}\big) = \frac{|\mathbf{a}^{(t)} \cap \mathbf{a}^{(t-1)}|}{|\mathbf{a}^{(t)} \cup \mathbf{a}^{(t-1)}|},
\]
where the numerator counts the number of edges present in both networks, and the denominator counts the number of edges present in at least one of the two networks. This yields a value between zero and one, with one indicating identical edge sets and zero indicating no common edges.

\begin{figure}[ht]
    \centering
    \includegraphics[width=0.8\textwidth]{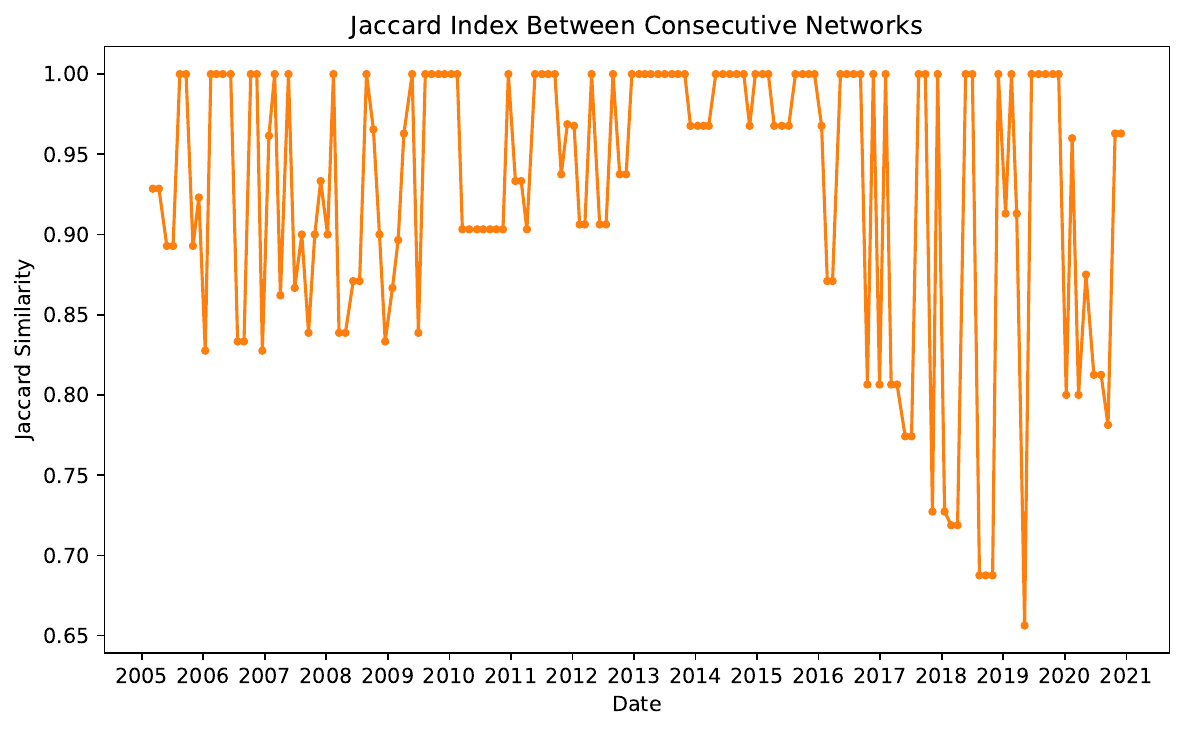}
    \caption{Jaccard similarity index between consecutive GL networks. Values closer to one indicate higher structural persistence.}
    \label{fig:jaccard}
\end{figure}

As shown in Figure~\ref{fig:jaccard}, the Jaccard index remains close to one for much of the sample, indicating that many edges persist across consecutive refit windows. From around 2016 onwards, however, the index becomes more erratic, with frequent dips below 0.8. This coincides with the period of greater edge-count variability and suggests that the estimated network structure is less stable in later years.

This instability can be interpreted in two ways. On one hand, it may reflect heightened sensitivity of the graphical lasso to noise, reducing the robustness of the estimated networks for forecasting. On the other hand, it may be capturing genuine time variation in cross-market linkages, in which case the shifting networks could be seen as appropriately adapting to changing market regimes. A possible explanation for the weaker forecasting performance of GL-based models is that, while they can adapt, the monthly refitting on a three-year rolling window may be too slow to fully capture more abrupt structural shifts, leaving parts of the estimated network stale between refits.

Summary of results:

\begin{itemize}
    \item The best-performing model found by QLIKE is a \textbf{local GNAR-HAR} with an FC network and no exogenous variables.

    \item The lowest MSE is achieved by a \textbf{standard GNAR-HARX} model with implied volatility as the sole exogenous input, also using an FC network.

    \item \textbf{FC networks} generally outperformed GL networks.

    \item \textbf{Overnight returns} and \textbf{implied volatility} appear frequently in the top-performing GNAR-HARX configurations.

    \item Standard univariate models (HAR, HARX) were consistently outperformed by GNAR-HAR(X) models, while often having more estimated parameters.

    \item Estimated model coefficients, the autoregressive ($\alpha$), network spillover ($\beta$), and exogenous ($\lambda$) terms, showed meaningful variation over time.

    \item Analysis of GL network structure revealed stability in earlier years, but increasing variance and reduced structural persistence from 2016 onward.

    \item Residual analysis indicated generally good model fit, but some underestimation during extreme volatility events, with evidence of skewness and heavy tails.
\end{itemize}

\section{Discussion and Future Work}
\label{sec:discussion}

\subsection{Discussion}

Our results demonstrate that GNAR-HAR(X) models outperform univariate HAR/HARX benchmarks, even when there are considerably fewer parameters to estimate. The top-performing specification as judged by QLIKE was a local GNAR-HAR model with an FC network and no exogenous variables, while the best model by MSE was a standard GNAR-HARX model using IV. Notably, the univariate HARX models, while competitive when judged by QLIKE, show greater sensitivity to very large spikes in implied volatility, leading to occasional extreme forecast errors and higher MSE.

The consistent underperformance of GL networks, relative to FC networks, highlights an important trade-off. Although the GL networks are refitted dynamically to adapt to changing market conditions, this adaptivity came at the cost of stability. Early in the sample the estimated networks were relatively persistent, but from 2016 onward they became more volatile, with sharper drops in edge count and more frequent dips in Jaccard similarity of consecutive networks. This instability likely reflects both estimation noise and genuine structural change, but in either case it reduced forecasting accuracy. In this setting, the variance costs of adaptivity appear to outweigh its potential benefits, making the simpler FC network more effective.

Analysing model coefficient estimates over time for the best global GNAR-HARX model (FC network with IV) offers further insight into model behaviour. Autoregressive coefficients ($\alpha$) remained stable, with mid-horizon (weekly) lags dominant. Network spillovers ($\beta$), however, were less stable, with more variation evident following crisis periods such as 2008 and 2020. The exogenous IV coefficient ($\lambda_{\text{IV}}$) was generally small but exhibited distinct spikes during market stress, consistent with its interpretation as a forward-looking risk indicator.

Residual analysis of the best global GNAR-HARX model confirmed a good overall fit but revealed slight underestimation during extreme volatility events. The residuals showed mild positive skew and excess kurtosis in the $\log{RV}$ space, suggesting that alternative error models with heavier tails (e.g., Student-$t$) may be more appropriate.

\subsection{Limitations and Future Work}

While the GNAR-HARX framework demonstrates promising empirical performance, several limitations remain that suggest directions for future research.

First, we do not compare directly against multivariate volatility models without a network structure, such as VAR-HAR \citep{bubak_volatility_2011, soucek_realized_2013}. These could serve as a natural benchmark to isolate the added value of incorporating a network structure. Given the large number of models already evaluated, such comparisons were left for future work, but would clarify whether the GNAR structure offers unique benefits over more standard multivariate alternatives. Future work might focus on a narrower comparison between the best GNAR-type models identified here and alternative multivariate approaches.

Second, the assumption of a known or correctly estimated network structure is strong. While we implemented graphical lasso to construct networks from the estimated time-varying sparse precision matrices, the resulting networks appeared less stable in later years of the sample, potentially undermining the forecasting performance of GL-based models. Alternative strategies may perform better. For example, \citet{nason_forecasting_2025} report success using randomly generated networks in the RaGNAR model for UK inflation forecasting. Such findings suggest that stochastic or ensemble-based network designs could offer a promising direction for future research.

Third, our analysis was limited to ten international stock indices with available high-frequency and implied volatility data. This restriction was driven by data availability and the need for reliable implied volatility series. Expanding to a broader cross-section of markets, or applying the framework to individual stocks, where IV could be constructed from options data, would allow for a broader test of model scalability and generalisability.

Fourth, model estimation relied on ordinary least squares (OLS), which may become unreliable as the number of predictors grows with the number of nodes or exogenous variables. This is especially true in local GNAR-HAR(X) models. Regularised regression methods, such as ridge or lasso, could improve estimation stability and predictive accuracy. Preliminary results with ridge regression on simulated data from local models support this idea, and future work could explore penalised or Bayesian estimation.

Fifth, the residual analysis suggests departures from normality, with heavier tails and occasional large outliers, a common feature in financial volatility data. It would be natural to consider heavier-tailed error distributions, such as the Student-$t$, to account for occasional large outliers more effectively. Some work in this direction has already been undertaken for GNAR-type models \citep{olawale_olanrewaju_vulgarized_2023}. Another promising extension is to model continuous and jump components of volatility separately, for example by combining GNAR-type dependence structures with jump-robust realised variance estimators \citep{andersen_roughing_2007}.

Finally, while all models in this analysis were evaluated using one-step-ahead forecasts, many practical applications require multi-step horizons. Extending the framework to multi-horizon forecasting, while also enriching the lag structure of exogenous variables beyond the single daily lag considered here, would provide a more comprehensive assessment of model performance and may enhance forecast accuracy.

To summarise, while the GNAR-HARX framework offers a flexible and interpretable structure for multivariate volatility forecasting, there remains considerable scope for methodological refinement and broader empirical validation.

\newpage
\section{Endmatter} \label{sec:endmatter}

\begin{itemize}
    \item \textbf{Data and reproducibility:} The code and data required to reproduce the results of this study are available in a private GitHub repository at \url{https://github.com/tomonuallain/msc-project}. Access can be granted on request via email (\texttt{tso24@ic.ac.uk} or \texttt{tomonuallain@gmail.com}).

    \item \textbf{Use of AI tools:} Conversational AI tools (ChatGPT) were occasionally used to assist with rephrasing or shortening sentences. All substantive content, analysis, and interpretation are my own.

\end{itemize}

%
%
%
%
%
%

%
%

\clearpage

\bibliographystyle{apalike2}
\bibliography{references}

\clearpage

%
%
%
%

\pagenumbering{arabic}
\renewcommand*{\thepage}{Supplementary Material Page \arabic{page}}
\renewcommand{\thesection}{\Alph{section}}
\appendix
\pagebreak
\section{Derivation of Stationarity Constraint}
\label{app:stationarity}

Assuming stationarity of the exogenous processes $\{X_{h,i,t}\}$, we follow Appendix A.1 of \cite{tapia_costa_higher_2025}, where GNAR-HAR is formulated as a GNAR(22, $\mathbf{r}$) model. The HAR coefficients correspond to:

\begin{align*}
    \alpha_{i,1} &= \alpha_d, \quad 
    \alpha_{i,j} = \alpha_w / 4 \text{ for } j=2,\dots,5, \quad 
    \alpha_{i,j} = \alpha_m / 17 \text{ for } j=6,\dots,22, \\
    \beta_{1,r} &= \beta_{d,r}, \quad 
    \beta_{j,r} = \beta_{w,r}/4 \text{ for } j=2,\dots,5, \quad 
    \beta_{j,r} = \beta_{m,r}/17 \text{ for } j=6,\dots,22.
\end{align*}

Here the per–lag maximum neighbourhood stages are
$s_1 = r_d$, $s_j = r_w$ for $j=2,\dots,5$, and $s_j = r_m$ for $j=6,\dots,22$.

For a GNARX($p, \mathbf{s}, \mathbf{p}'$) process with stationary exogenous inputs, the sufficient condition for stationarity from \cite{nason_quantifying_2022} is:

\[
\sum_{j=1}^{p} \left( |\alpha_{i,j}| + \sum_{s=1}^{s_j} |\beta_{j,s}| \right) < 1 \quad \forall i.
\]

Substituting in the HAR-to-GNARX parameter mappings, we obtain:

\[
\begin{aligned}
\sum_{j=1}^{22} \left( |\alpha_{i,j}| + \sum_{r=1}^{s_j} |\beta_{j,r}| \right) 
&= |\alpha_d| + \sum_{r=1}^{r_d} |\beta_{d,r}| \\
&+ \sum_{j=2}^{5} \left( \frac{|\alpha_w|}{4} + \sum_{r=1}^{r_w} \frac{|\beta_{w,r}|}{4} \right) \\
&+ \sum_{j=6}^{22} \left( \frac{|\alpha_m|}{17} + \sum_{r=1}^{r_m} \frac{|\beta_{m,r}|}{17} \right) \\
&= |\alpha_d| + |\alpha_w| + |\alpha_m| + \sum_{r=1}^{r_d} |\beta_{d,r}| + \sum_{r=1}^{r_w} |\beta_{w,r}| + \sum_{r=1}^{r_m} |\beta_{m,r}|.
\end{aligned}
\]

This yields the constraint stated in the main text.

\section{Jensen’s Inequality Adjustment}
\label{app:jensen}
In this project, models are estimated on the log of realised variance, 
so that forecasts are obtained for 
$Y_{i,t} = \log RV_{i,t}$. 

If $X \sim \mathcal{N}(\mu,\sigma^2)$, then
\[
\mathbb{E}[e^X] 
= \int_{-\infty}^{\infty} e^x \, \frac{1}{\sqrt{2\pi\sigma^2}}
\exp\!\left(-\frac{(x-\mu)^2}{2\sigma^2}\right) dx.
\]
Completing the square inside the exponential gives
\[
\mathbb{E}[e^X] = \exp\!\left(\mu + \tfrac{1}{2}\sigma^2\right).
\]
Applying this conditionally with $X=\widehat{Y}_{i,t}+u_{i,t}$, where $u_{i,t}\mid \mathcal F_{t-1}\sim\mathcal N(0,\sigma_{i,t}^2)$, yields
\[\mathbb{E}[RV_{i,t}\mid \mathcal F_{t-1}] = \exp\!\left(\widehat{Y}_{i,t} + \tfrac{1}{2}\sigma_{i,t}^2\right).\]
which is the corrected back-transformation in \eqref{eq:jensen_adjustment}.

\newpage
\section{Graphical Lasso Network Example}
\label{app:gl_network}
\begin{figure}[ht]
    \centering
    \includegraphics[width=0.8\textwidth]{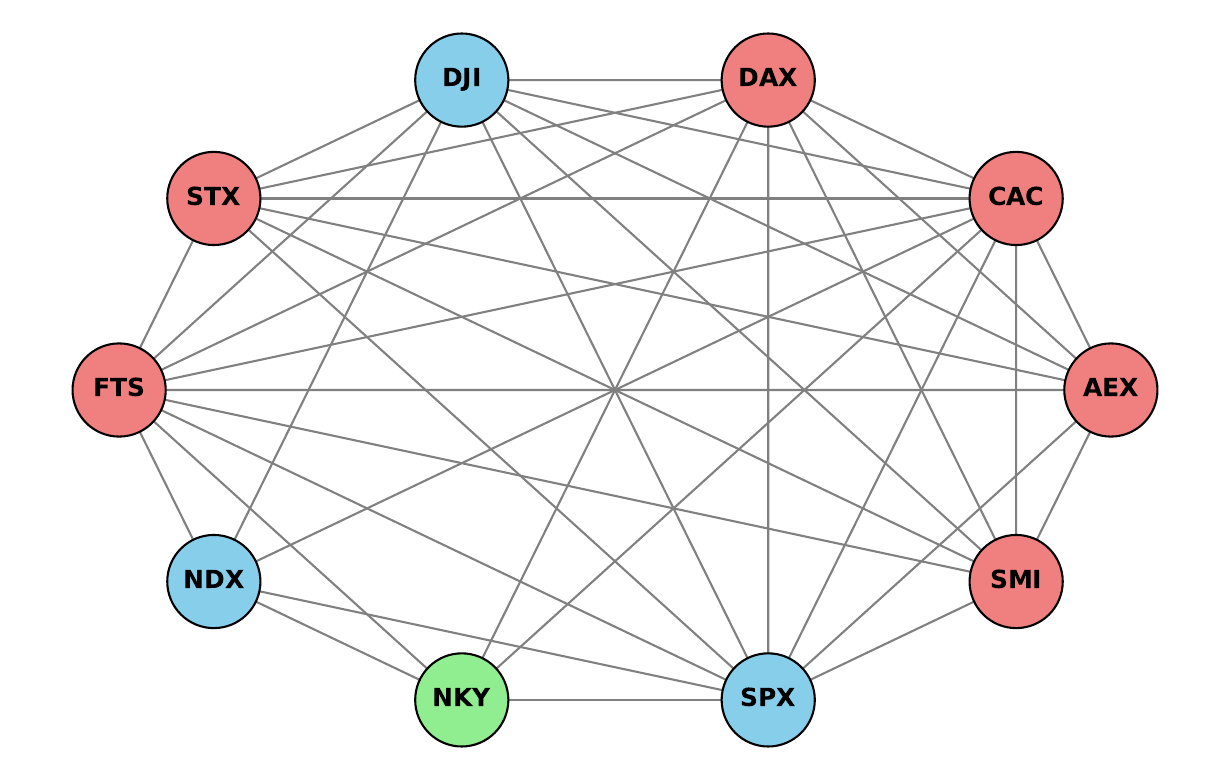}
    \caption{Example of a graphical lasso network, with nodes coloured by region.}
    \label{fig:gl_network}
\end{figure}

\newpage
\section{Further Residual Analysis}
\label{app:residual-other}

\begin{figure}[h]
    \centering
    \includegraphics[width=0.8\textwidth]{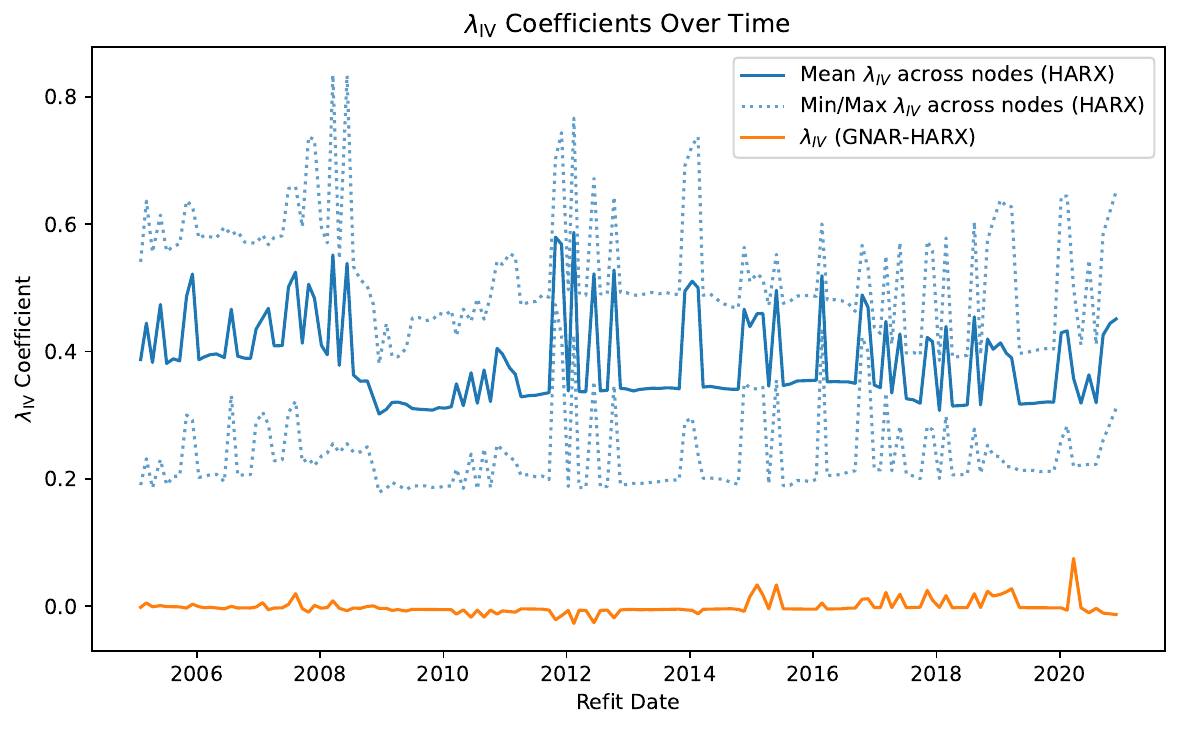}
    \caption{Estimated IV coefficients ($\lambda_{\text{IV}}$) over refits for HARX (node-wise mean and range) and global GNAR-HARX. HARX shows larger and more variable loadings on IV, whereas GNAR-HARX loadings are near zero and stable.}
    \label{fig:iv_harx_vs_gnarharx}
\end{figure}

\begin{figure}[H]
    \centering
    \includegraphics[width=0.8\textwidth]{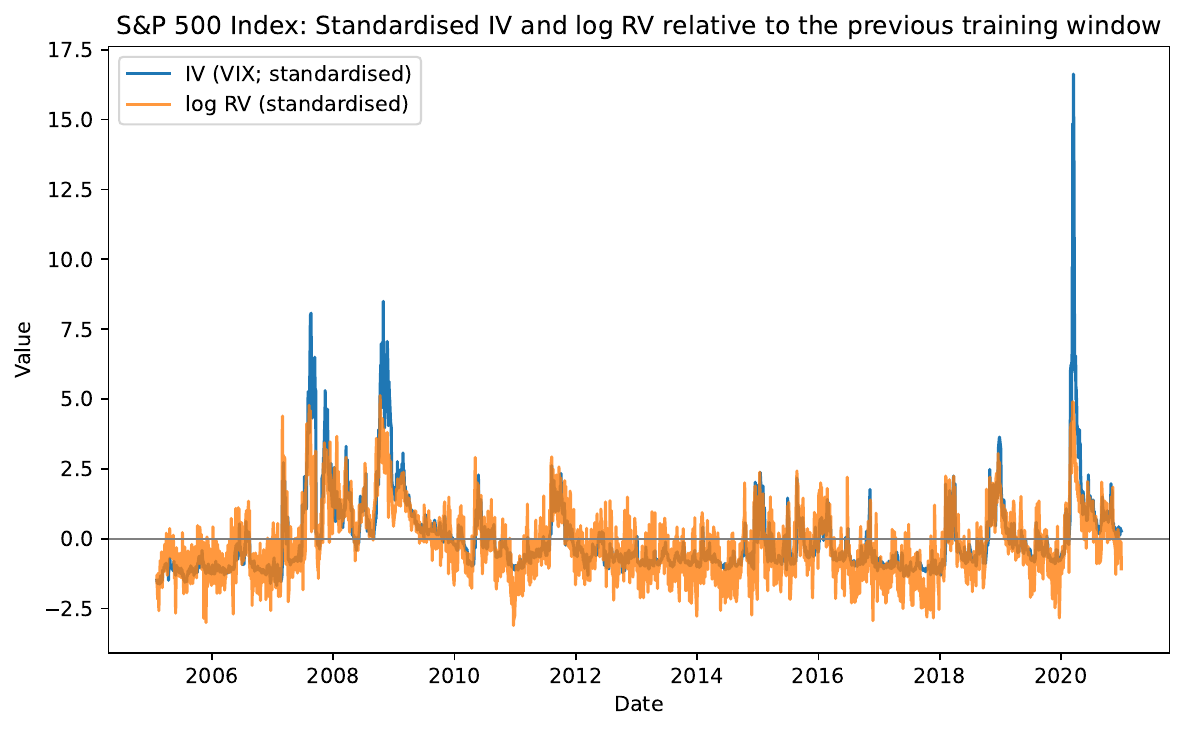}
    \caption{S\&P 500 (SPX) Index: IV (VIX) vs $\log RV$, both standardised relative to the previous refit window. The IV spike in March 2020 helps explain HARX(IV) MSE instability.}
    \label{fig:iv_vs_rv}
\end{figure}

\newpage
\begin{figure}[ht]
    \centering
    \includegraphics[width=\textwidth]{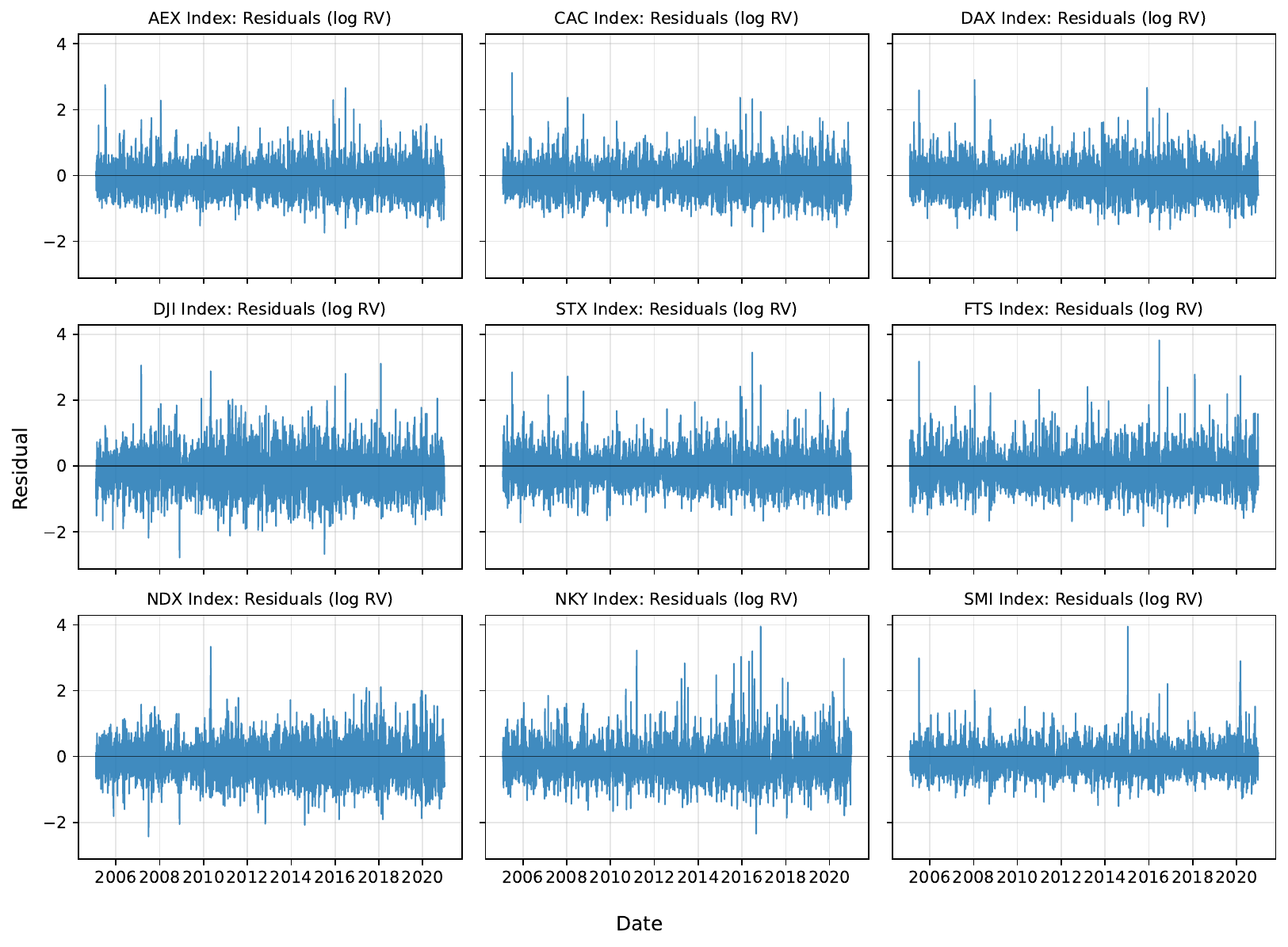}
    \caption{Residuals (log-realised variance) over time for the other nine indices under the best global GNAR-HARX model (FC network with IV). Residuals are generally centred around zero, with occasional large deviations.}
    \label{fig:residuals-all}
\end{figure}

\begin{figure}[ht]
    \centering
    \includegraphics[width=\textwidth]{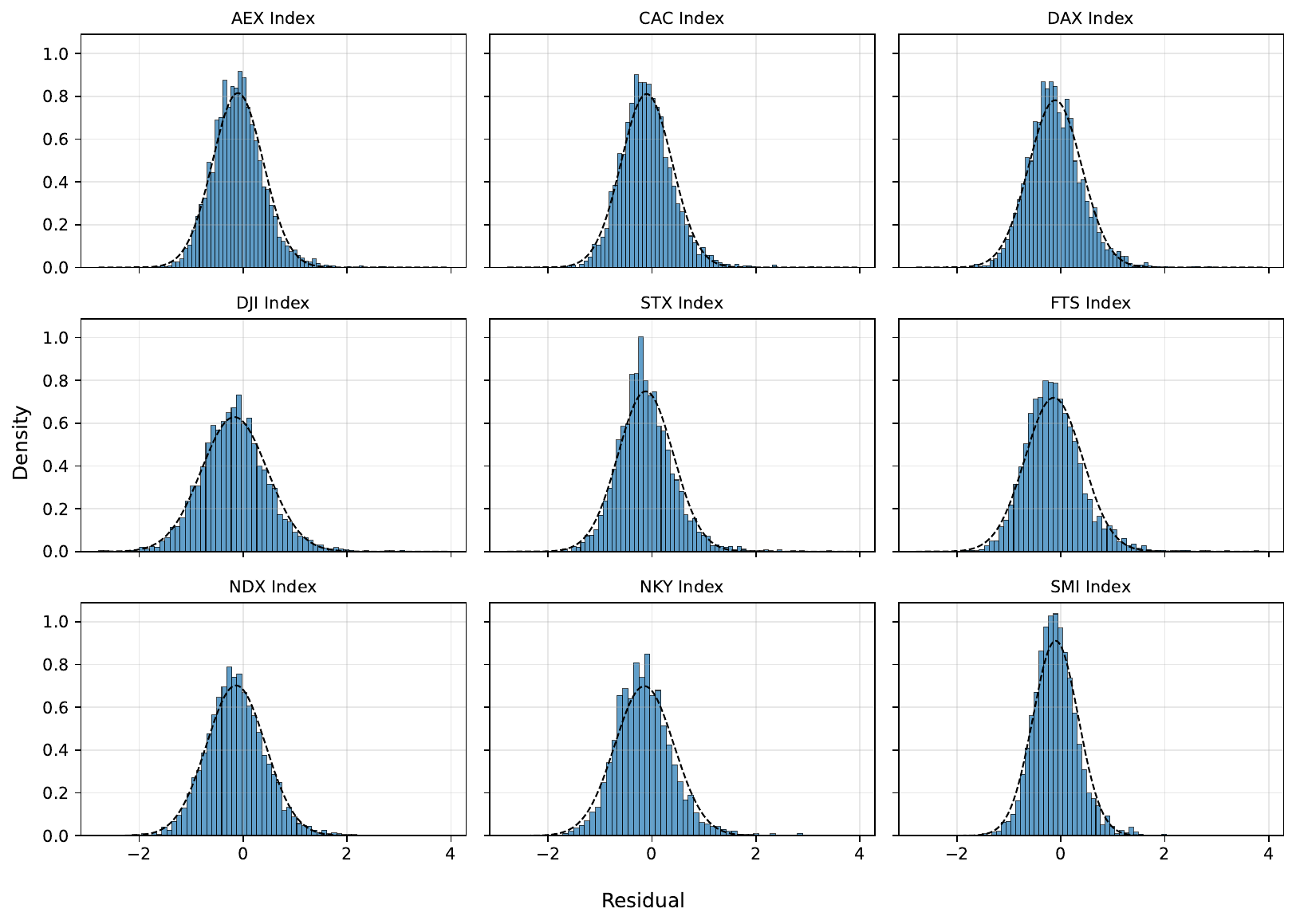}
    \caption{Distribution of residuals (log-realised variance) for the other nine indices under the best global GNAR-HARX model (FC network with IV). Histograms are overlaid with a normal density fit. Distributions are broadly symmetric but exhibit heavier tails than normality, consistent with the SPX case.}
    \label{fig:residuals-all-dist}
\end{figure}

\end{document}